**Sizing Battery Energy Storage and PV System in an Extreme Fast Charging Station Considering Uncertainties and Battery Degradation**

Waqas ur Rehman, Rui Bo*, Hossein Mehdipourpicha, Jonathan Kimball

Department of Electrical and Computer Engineering, Missouri University of Science and Technology, Rolla, MO 65409 USA.

*(\*) Corresponding author e-mail:* rbo@mst.edu.

**Abstract**

This paper presents mixed integer linear programming (MILP) formulations to obtain optimal sizing for a battery energy storage system (BESS) and solar generation system in an extreme fast charging station (XFCS) to reduce the annualized total cost. The proposed model characterizes a typical year with eight representative scenarios and obtains the optimal energy management for the station and BESS operation to exploit the energy arbitrage for each scenario. Contrasting extant literature, this paper proposes a constant power constant voltage (CPCV) based improved probabilistic approach to model the XFCS charging demand for weekdays and weekends. This paper also accounts for the monthly and annual demand charges based on realistic utility tariffs. Furthermore, BESS life degradation is considered in the model to ensure no replacement is needed during the considered planning horizon. Different from the literature, this paper offers pragmatic MILP formulations to tally BESS charge/discharge cycles using the cumulative charge/discharge energy concept. McCormick relaxations and the Big-M method are utilized to relax the bi-linear terms in the BESS operational constraints. Finally, a robust optimization-based MILP model is proposed and leveraged to account for uncertainties in electricity price, solar generation, and XFCS demand. Case studies were performed to signify the efficacy of the proposed formulations.

**Nomenclature**

*Indices*

| | |
|---|---|
| $i$ | Index of all EVs considered for XFCS demand modeling |
| $j$ | Index of scenarios to consider seasonal variations of the input data for both weekdays and weekends |
| $k$ | Index for power imported from power distribution network (PDN) averaged over $\xi$-mins window |
| $t$ | Index for time |
| $\varsigma$ | Index for BESS cycle life curve segment |
| $\mathcal{s}$ | Index of considered seasons in a year (i.e., winter, spring, summer, and fall) |
| $a$ | Index for EVs driven mileage |

*Parameters*

| | |
|---|---|
| $CE_{BESS}^{cap}$ | Annualized capital cost of BESS energy rating [\$/kWh] |
| $CE_{BESS}^{inst}$ | Annualized BESS installation cost [\$/kWh] |
| $CP_{BESS}^{cap}$ | Annualized capital cost of BESS power rating [\$/kW] |
| $C_{BESS}^{O\&M}$ | Annualized BESS O&M cost [\$/kW] |
| $CP_{PV}^{cap}$ | Annualized capital cost of PV system power rating [\$/kW] |
| $C_{PV}^{O\&M}$ | Annualized PV system O&M cost [\$/kW] |
| $\underline{C_{BESS}}/\overline{C_{BESS}}$ | Lower/upper limit of BESS energy capacity [kWh] |
| $\mathfrak{C}_i^{EV}$ | Battery capacity of EV $i$ |
| $\underline{DoD}/\overline{DoD}$ | Lower/upper limit of BESS depth of discharge [%] |
| $\widehat{E_{XFCS,j}}(t)$ | XFCS daily charging demand forecast for scenario $j$ at time $t$ [kWh] |
| $E_{XFCS,j}^d(t)$ | Uncertain XFCS daily charging demand for scenario $j$ at time $t$ [kWh] |
| $E_{XFCS}^{max}$ | XFCS maximum energy demand [kW] |
| $E_i^{cpm}$ | Per mile energy consumption of EV $i$ [kWh/mi] |
| $E_{i^*}^{EV}(t)$ | Energy needed to charge EV $i^*$ at time $t$ |
| $E_{WD}(t)/E_{WND}(t)$ | Aggregated energy demand profile of the XFCS for a typical weekday and weekend day at a time $t$ |
| $\mathcal{F}(\mathcal{Q}_a)$ | Probability of driving $\mathcal{Q}_a$ average mileage [%] |
| $h(q)$ | Exponential distribution representing daily driven mileage |
| $IR_{G,j}(t)$ | Global irradiance data on a fixed plane for scenario $j$ at time $t$ [W/m$^2$] |
| $\mathfrak{T}^*$ | Total number of EVs that got recharged at the station |
| $\mathcal{K}$ | Total number of $\xi$-mins time intervals in a day |
| $\mathcal{L}$ | Project lifetime |
| $\mathbb{M}$ | A large number used in Big-M method to avoid simultaneous charging/discharging of BESS |
| $NOCT$ | Normal operating cell temperature [°C] |
| $\mathcal{N}(\mu, \sigma^2)$ | Normal probability distribution with mean μ and variance $\sigma^2$ |
| $\overline{PV_{G,j}}(t)$ | Estimate of per unit PV system generation power for scenario $j$ at time $t$ |
| $PV_{G,j}^p(t)$ | Uncertain per unit PV system generation power for scenario $j$ at time $t$ |
| $P^{XFCS}$ | Rated charging power of single charging power at the XFCS |
| $Q_a$ | Average driven mileage [miles] |
| $\widehat{Q_a}$ | Average driven mileage for which EV's SoC does not reach to its $SOC_i^{thr}$ [miles] |
| $Q_a^*$ | Average driven mileage for which EV's SoC reaches to its $SOC_i^{thr}$ [miles] |
| $Q_d, Q_{d+1}$ | Distance values used to discretize the daily driven mileage and obtain the average driven mileage ($Q_a$) in interval $[Q_d, Q_{d+1}]$ [miles] |
| $r$ | Number of charging ports inside the XFCS |
| $w$ | Number of waiting spots inside the XFCS |
| $SF^M$ | Factor to scale monthly costs to yearly cost |
| $SoC_{i,a}^{EV}(t)$ | State of charge of EV $i$ at time $t$ after driving total distance of $Q_a$ miles |
| $SoC_i^{thr}$ | SoC threshold for EV $i$ at which it should start recharging at the XFCS |
| $\widehat{SOC}_{i^*}^{arr}$ | Weighted average of the arrival SoC of EV $i^*$ arriving at the station after traveling $Q_a^*$ miles |
| $SOC_{i^*}^{target}$ | Desired target SoC for recharging EV $i^*$ |
| $\hat{t}_{i^*}^{arr}$ | Weighted average of the arrival time of EV $i^*$ arriving at the station after traveling $Q_a^*$ miles |
| $t_{i^*}^{st}$ | Start time to charge EV $i^*$ |
| $t_{i^*}^{ch}$ | Duration for which EV $i^*$ gets recharged till its battery reaches to $SOC_{i^*}^{final}$ |
| $\mathcal{T}$ | Total minutes in a day [min] |
| $T_{Amb,j}(t)$ | Daytime ambient temperature for scenario $j$ at time $t$ [°C] |
| $z$ | interest rate [%] |
| $\mathbb{S}$ | Total number of BESS cycle life curve segments |
| $\varpi$ | Coefficient of exponential distribution $h(q_i)$ |
| $\pi$ | BESS ramp rate limit [kWh/min] |
| $\underline{\gamma}/\overline{\gamma}$ | Minimum/maximum ratios of BESS energy and power rating |
| $\mathcal{J}$ | Total number of considered scenarios to represent a typical year |
| $\mathfrak{D}$ | Total number of days in a year |
| $\mathcal{S}$ | Total number of considered seasons in a year (i.e., winter, spring, summer, and fall) |
| $\xi$ | Averaging window for power import from the PDN [min] |
| $\underline{\Psi}/\overline{\Psi}$ | Lower/upper limit of BESS charge/discharge cycles |
| $\lambda_{ADC}/\lambda_{MDC}$ | Annual/monthly demand charges [\$/kW] |
| $\hat{\lambda}_{E,j}(t)$ | Hourly wholesale electricity market price forecast for scenario $j$ at time $t$ [\$/kWh] |
| $\eta_{AC-DC}$ | AC/DC conversion efficiency [%] |
| $\eta_{DC-DC}$ | DC/DC conversion efficiency [%] |
| $\eta_{ch}/\eta_{dch}$ | BESS charging/discharging efficiencies [%] |
| $\mathfrak{p}$ | Power temperature coefficient |
| $\rho(s)$ | Non-negative continuous weight variable associated with BESS cycle life curve segment $s$ |
| $\mathfrak{J}(DoD(s))$ | Number of BESS cycles associated with depth of discharge $DoD(s)$ |
| $\varphi^{EV}(t)$ | Probability of driving an EV at time $t$ |
| $\Phi, \varrho, \mathcal{E}$ | Maximum deviation from forecasted values of $\hat{\lambda}_{E,j}(t)$, $\widehat{E_{XFCS,j}}(t)$, $\widehat{PV_{G,j}}(t)$, respectively |
| $\Gamma^\lambda, \Gamma^D, \Gamma^{PV}$ | Uncertainty budgets for electricity market price forecast, XFCS charging demand forecast, and per unit PV generation forecast, respectively |
| $U^\lambda, U^D, U^{PV}$ | Uncertainty sets for electricity market price forecast, XFCS charging demand forecast, and per unit PV generation forecast, respectively |
| $\Delta t$ | Simulation time step [1/60 h] |

*Variables and functions*

| | |
|---|---|
| $CF_{ann}$ | Annualized cost factor |
| $C_{BESS}$ | BESS energy capacity [kWh] |
| $DoD$ | Depth of discharge of BESS |
| $DoD(\mathfrak{s})$ | DoD associated with BESS cycle life curve segment '$\mathfrak{s}$' |
| $E_{ch,j}(t)$ | Energy flow to charge the BESS for scenario $j$ at time $t$ [kWh] |
| $E_{dch,j}(t)$ | Energy flow due to BESS discharging for scenario $j$ at time $t$ [kWh] |
| $E_{BESS,j}(t)$ | BESS stored energy for scenario $j$ at time $t$ [kWh] |
| $P_{PV}$ | PV system rated power [kW] |
| $P_{pv,j}(t)$ | PV system power generation profile for scenario $j$ at time $t$ [kW] |
| $E_{g,j}^{AC}(t)$ | AC energy exchanged with the PDN for scenario $j$ at time $t$ [kWh] |
| $E_{g,j}^{+}(t)$ | AC energy imported from the PDN for scenario $j$ at time $t$ [kWh] |
| $E_{g,j}^{-}(t)$ | AC energy exported to the PDN for scenario $j$ at time $t$ [kWh] |
| $P_{g,avg,j}^{+}(k)$ | $\xi$-mins average AC power import for scenario $j$ and for $k^{th}$ interval [kW] |
| $P_{g,an}^{max}$ | Maximum $\xi$-mins average annual AC power import [kW] |
| $P_{g,mo,\mathfrak{s}}^{max}$ | Maximum $\xi$-mins average monthly AC power import for season $\mathfrak{s}$ [kW] |
| $P_{g,daily,j}^{max}$ | Maximum $\xi$-mins average daily AC power import for scenario $j$ [kW] |
| $P_{BESS}^{rated}$ | Rated BESS charge/discharge power [kW] |
| $SoC_{BESS,j}(t)$ | BESS state of charge for scenario $j$ at time $t$ [%] |
| $\lambda_{E,j}^{\ell}(t)$ | Uncertain wholesale electricity market price for scenario $j$ at time $t$ [\$/kWh] |
| $E_{XFCS,j}^{d}(t)$ | Uncertain XFCS daily average charging demand forecast for scenario $j$ at time $t$ [kWh] |
| $PV_{G,j}^{\wp}(t)$ | Uncertain yearly per unit PV system generation power for scenario $j$ at time $t$ |
| $\alpha, \beta_j(t)$ | Dual variables of the original DO model |
| $\varsigma_j(t)$ | Auxiliary variable used to achieve linearized expressions in the RO model |
| $\Psi$ | Annual BESS charge/discharge cycles |
| $\mathfrak{J}^a(DoD)$ | Allowed number of BESS cycles using piece-wise linear approximation of BESS cycle life curve |

*Binary variables*

| | |
|---|---|
| $u_{1,j}$ | Binary variable used in Big-M method to avoid simultaneous grid energy import/exports for scenario $j$ at time $t$ |

$u_{2,j}$         Binary variable used in Big-M method to avoid simultaneous charging/discharging of BESS for scenario $j$ at time $t$

## 1 Introduction

The emission of greenhouse gases (GHG) from fossil fuel energy resources elevated concerns about climate change and global warming. Global temperature variation due to human engagements is estimated to be 1°C [1]. Road transportation using internal combustion engine vehicles accounts for over 70% of the GHG emissions [2]. Transportation electrification can prove pivotal in reducing the effects of GHG emissions and carbon footprints on the environment. However, extended charging time and the *range anxiety* associated with electric vehicles (EVs) is still a major challenge to surmount. These limitations affect the EV adoption by city drivers (due to extended charging times) and highway drivers (due to range anxiety). In the recent few years, efforts were made to lower recharge time by developing and installing direct current (DC) fast charging stations—a category of fast chargers that recharge EVs by supplying the DC power directly and have charging time of more than 1 h with a power level of up to 50 kW for a 200-mi range [3-8]—in public places, but the time is still not comparable to refueling time at a conventional gasoline station. The extreme fast charging technology, conversely, is capable of recharging EVs in 10 minutes—which is comparable to refueling gasoline vehicles—with a peak power level of 350 kW for the 200-mi range [4, 9-11]. Per [4, 10, 12, 13], the charging stations with rated charging power of 350 kW and above are categorized as extreme fast charging stations. Therefore, the deployment of extreme fast charging stations (XFCS) in urban areas, rural areas, and on highways can prove essential for the proliferation of EVs and electrified transportation.

      Extreme fast charging of EVs may cause various issues in power quality of the host power grid, including power swings of $\pm 500$ kW [14], subsequent voltage sags and swells, and increased network peak power demands due to the large-scale and intermittent charging demand [15, 16]. If the XFC charging demand is not managed prudently, the increased daily peak demand and a shift in daily peak due to EV charging may cause transformer and feeder overload, accelerating transformer aging, and increasing power losses [11, 17]. Consequently, grid reinforcement and expansion planning would become essential to meet the increasing charging demand and enable XFC, especially on the distribution level where XFCS are directly connected. A battery energy storage system (BESS) can act as a *power buffer* to mitigate the transient impact of the extreme fast charging on the power distribution network (PDN) power quality [18]. It can also act as an *energy buffer* to charge energy during low-price hours and discharge it during high-price hours to earn revenue, thereby not only reducing the overall operational cost of the XFCS but also avoiding huge investment costs on the grid reinforcement and expansion planning [16]. This process is called energy or price arbitrage [19, 20]. Renewable generation resources, such as PV systems, can generate low-cost energy compared to the energy purchased from the PDN to fulfill the local demand; excess energy can be sold to the power grid [21, 22]. Therefore, the

installation of a PV system in an XFCS can yield extra savings in the XFCS operation cost. In addition, the installation of a PV system and a storage system can reduce the PDN peak demand increment caused by charging station operation. Currently, the number of EV charging stations that rely only on the electric grid to recharge EVs is higher than those that are assisted by renewable resources and BESS. Nevertheless, there have been rapid advancements in the proliferation of the latter in recent few years: Tesla Inc. is spearheading the efforts to electrify the road transportation system, and as of year-end 2021 it has installed over 30,000 next-generation V3 Superchargers globally—that are capable of recharging EVs in 15-mins for 200 miles range with a max recharge rate of 250 kW—and some of which are assisted by the solar generation and battery storage systems [23-25]. In addition, Tesla also plans to power all of its superchargers with renewable energy and battery storage in the near future [26, 27].

## 1.1  Related Work

Planning of privately owned EV charging stations has been attempted in various studies in the literature. This section presents a comprehensive review of the extant literature related to charging station components' sizing, research gap identification, and this study's field contributions.

In [28], Ding et al. presented a mixed integer linear programming (MILP) model to assess the capacity of the Li-ion based BESS to (i) reduce peak power import from the PDN, (ii) downsize transformer and feeder capacity, (iii) exploit energy buffering for energy arbitrage, and (iv) alleviate the charging demand variance in an electric bus (EB) charging station. Negarestani et al. [29] proposed a MILP model to obtain the energy capacity of the flywheel storage for energy arbitrage in a fast charging station (FCS). Salapić et al. [30] also proposed a MILP model to find the BESS capacity size to reduce the FCS operational cost and stress on the PDN. Moreover, net present value (NPV) was employed to find the cost of investing in the BESS. Bryden et al. [31] approached sizing of the Li-ion based BESS to reduce the power rating of the connection to the power network, thus mitigating the necessity for potential grid infrastructure reinforcement. Another objective was to reduce the mean waiting time for the EVs arriving at the FCS. The proposed method was comprised of two stages: first, the number of charging ports was determined, and in the second stage, the optimal energy capacity of the BESS was obtained. The relationship between BESS capacity and users' average waiting time was explored for sizing purposes. Monte Carlo Simulations (MCS) were used to get the optimal capacity of the BESS based on reasonable users' average waiting time at the charging station. However, in this work, the operational characteristics and investment cost of the BESS were not considered. MCSs were used in [32] to obtain optimal sizing for the storage system and grid-tie converter, such that the grid-tie converter was designed to provide the average power demand and storage system to provide for peak power demands at the station. Storage system choice was made after the sizing problem was solved. The ultracapacitor-based storage system was found to be the best choice based on the requisite power-capacity combination obtained from simulations. In [33], an optimization model was developed that took into account the uncertainty of EVs arrival times, worst-case SoCs of EVs arriving at the station, and power level to recharge EVs. Objectives of the proposed model were to minimize the annual operational cost (AOC) of the charging station, the annual penalty cost associated with charging demand during peak periods, and the investment cost of the Li-ion based BESS. The output of the model was the simultaneous sizing of the BESS and converters, thereby avoiding the over- or under-sizing of the charging station components.

A quantitative stochastic model was used in [34] to determine the sizing of local BESS by analyzing its relationship with the quality of service to customers, expressed as *customer blocking probability*. Corchero et al. [35] proposed an optimization model to provide more charging power to EVs than permitted by grid connection and minimize the operational cost of the EV charging energy, investment cost, and operation and maintenance (O&M) cost of the charging station components. The output of the proposed model had optimal capacity ratings of BESS, optimal power ratings of the grid-tie converter, and optimal power flow between the grid and the FCS. A sensitivity-based analysis was performed in [36] to evaluate the impact of BESS (Li-titanate battery) capacity on reduction of AOC in the EB charging station, and the optimal value of the BESS kWh was achieved when reduction in AOC converged to a constant value. In this study, AOC was comprised of equivalent capacity charges for grid integration of the station, energy purchased from the grid, and the life expenditure cost of the BESS. Tan et al. [37] formulated a two-stage stochastic optimization problem to decide on sizing the BESS, transformer, and grid connection with regards to investment. Objectives of the proposed model were to (i) consider the participation of BESS in the electricity market and ancillary services simultaneously, and (ii) to minimize the total expected investment and annual operational cost of the FCS. A planning method was proposed in [38] to study the integration of FCS with the power grid to (i) minimize the BESS life cycle cost, installation cost, and grid connection cost of the FCS and (ii) mitigate the need for potential grid infrastructure reinforcement. A detailed cost-benefit analysis was performed to assess the viability of using BESS within the FCS by studying its life cycle and replacement cost. Finding the optimal BESS size posed a trade-off between the EVs charging demands and the grid constraints. Furthermore, this work introduced two BESSs concepts within the FCS for achieving partial decoupling between stations and the grid. A review of the literature, presented in [28-38] revealed that potential applications of renewable energy resources (RERs) and their optimal sizing were not investigated. Furthermore, life degradation considerations regarding the energy storage system—for instance, optimal depth of discharge (DoD), the allowable number of charge/discharge cycles, and calendric lifetime of the storage system—were not considered. A mixed integer non-linear programming (MINLP) model was proposed in [39] to optimally site and size an FCS to minimize the costs associated with station's development and loss of gird and electric vehicle energy. Nonetheless, applications of the energy storage system and RERs were not considered in this work.

Applications of RERs were investigated in [2, 40-44] and the charging station's planning problem was solved. A non-linear integer programming (NLIP) problem was formulated in [40] and solved using a search-based algorithm to find the optimum solar generation size and the energy storage system rating in a solar-powered off-grid charging station. A multi-dimensional discrete-time 3-D Markov chain model was used to incorporate the stochastic nature of the PV generation. Moreover, a queuing model was used to consider the randomness associated with EVs' arrivals at the station. However, operational and technical constraints of ESS were not considered. Furthermore, ESS degradation considerations were overlooked. In [41], Gunter et al. proposed a methodology for a charging station design that was integrated with RER and storage system, and it proved there were monetary benefits of having solar-generated power in the charging station. Hafez et al. [42] approached the charging station design problem by considering the lifecycle cost reduction and environmental emissions. Still, degradation of the storage system was not considered in either study. An MILP-based optimization model was proposed in [2] to optimally obtain the type and sizing of ESS and renewable sources for integrating with FCS. ESS degradation was considered in this work. However, to the best of our knowledge, for the planning of privately owned charging stations, the existing literature either completely ignored important data uncertainties—as associated with the charging station energy demand, renewable generation, and electricity market price for the energy acquired from the power grid—or used simple probabilistic methods to account for only some of the uncertainties. A bi-level robust optimization approach was proposed in [43, 44] it considered uncertainties in the electricity market prices, renewable generation, and EV users' behaviors. However, EV users' travelling behaviors and charging order were not included. Charging station demand modeling and related uncertainties were also not considered. In addition, the BESS degradation considerations and reduction of demand charges were ignored.

Accurate data prediction is a challenging task because planning input data are subject to uncertainties coupled with the charging demands, electricity market prices, and renewable generation levels. There are two common approaches to modeling uncertainties in planning problems: stochastic optimization (SO) modeling and robust optimization (RO) modeling. The SO approach needs accurate probability distribution of uncertainties to construct a large number of scenarios for precise characterization of the uncertain data; consequently, the planning model may be computationally expensive to solve and even intractable for large-scale systems [22, 45-48]. Moreover, obtaining the accurate probability distributions of long-term uncertainties, such as electricity market prices, EVs ownership data, EV drivers' traveling behaviors, and charging station demand, years into the future may be challenging. If uncertainties are not prudently accounted for, they may lead to significant errors in the sizing of the charging station components. By contrast, the RO approach only needs limited uncertain data information, in the form of uncertainty confidence bounds, and does not suffer from issues associated with the SO approach. Additionally, RO is computationally inexpensive compared to SO. Thus, RO methods are well suited for solving the planning problem of charging stations due to the involvement of the aforementioned long-term uncertainties [43-45]. In literature, RO was utilized for modeling and solving planning problems [22, 43, 44, 46, 47, 49-52] and scheduling/operation problems [48, 53-60]. However, the RO approach is conservative, and planners typically choose a trade-off between economy and robustness against uncertainties in practical applications [46]. Hence, it is worthwhile to study ways to lower the conservativeness of the results obtained from RO methods.

This work proposes a novel mathematical model for the problem of sizing the battery energy storage system and PV system in an XFCS by considering the application of BESS energy arbitrage, monthly and annual demand charges reduction, BESS life degradation, and uncertainties in the forecasted input parameters. We first proposed a model for the computation of XFCS charging demand; FCS demand modeling has been presented in the extant literature using fixed power-based EV recharge throughout the charging process, which is unrealistic and may introduce significant errors in the charging station components' planning. By contrast, this work uses realistic constant power constant voltage (CPCV) based extreme fast charging of EVs in the proposed demand modeling approach. One year was represented with eight scenarios and the proposed model was solved using a CPLEX solver. Additionally, sensitivity analyses were performed to understand how changing values of input parameters and different degrees of robustness against uncertainties in the input data influence the XFCS components' sizing and the station's total cost.

### 1.2 Contributions

In relation to the identified research gaps in existing literature, the contributions to the field are made by this study are summarized as follows:

1) We propose novel MILP formulations to find optimal power and energy ratings for a Li-ion based BESS, ratings for a PV system integrated with the station, and optimal energy management of the XFCS for each considered scenario. Unlike the extant literature on the planning of privately owned charging stations [2, 28-42, 61], this work also considers uncertainties—such as charging station demand, electricity market prices, and PV system generation—in the long-term forecast data, and it leverages the RO approach to model and solves the XFCS planning problem.

2) Compared to the probabilistic charging station demand modeling approaches in the literature [2, 19, 29], this work proposes an improved method to incorporate the realistic charging station demand characteristics by considering EV users' driving behaviors (such as the probability of vehicles' daily driven mileage, probability of daily trips for weekdays and weekends, etc.) and CPCV based extreme fast charging of EVs. Furthermore, this work incorporates the heterogeneous behavior of EV users in terms of starting and terminating the charging process using normal probability distributions of the starting SoC (i.e., $SoC^{thr}$) and desired target SoC (i.e., $SoC^{target}$) of EVs.

3) This work considers the reduction in monthly and annual demand charges associated with the XFCS maximum average power imported—monitoring rate used by utilities for calculating the demand charges is every 15 minutes (96 intervals/day) [62, 63]—from the PDN, based on realistic utility tariff [64], which were not considered in the extant literature. While the majority of the methods used in the literature incorporated hour-scale granularity, this study considers minute-scale granularity to compute the 15-min average power imported from the PDN and to cope with the rapid changes in the XFCS charging demand profile as it is capable of charging EVs in less than 10 minutes [4]. Thus, usage of minute-scale granularity is also critical for the correct power sizing of the BESS, in the case of XFCS sizing, as it depends on the information of actual charging profile and true peaks, which will be lost when hourly sampling is used.

4) This work accounts for the cycle-life degradation of the Li-ion based BESS when modeling the planning problem, which was neglected in the literature that examined charging station planning [28-38, 40-44]; thereby, this work ensures that the BESS will not be replaced during the lifetime of the project and consequently prevents extra investment cost that may incur otherwise. Furthermore, differing from the reported work related to BESS sizing [2, 65, 66], this work proposes novel and pragmatic linear formulations to accurately tally the number of BESS charge/discharge cycles using cumulative charging/discharging energy. The outcome of the proposed degradation model is the optimal DoD and the number of BESS charge/discharge cycles during the project's lifetime.

### 1.3 Paper Organization

This paper is structured as follows: Section 2 describes the XFCS demand model, Section 3 proposes a MILP-based deterministic optimization (DO) model, and Section 4 presents the RO-based optimization model for XFCS planning. Section 5 describes the case study and presents findings and discussion. Section 6 concludes the paper.

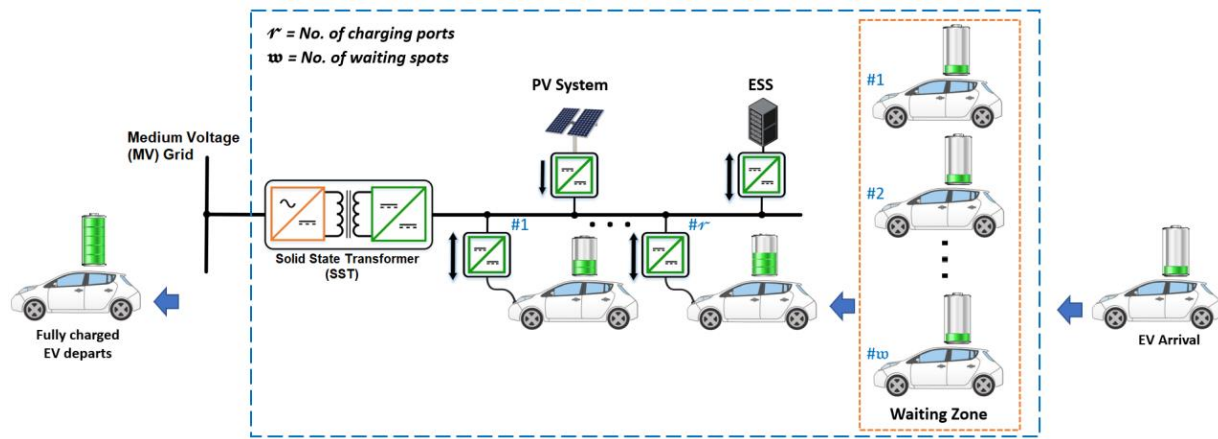

Fig. 1. XFCS schematic

## 2 XFCS Demand Modeling

This section presents the modeling of the XFCS charging demand profile. Fig. 1 presents a schematic of the XFCS that consists of BESS, PV system, and charging ports. The XFCS is connected to a medium voltage (MV) grid (e.g., 12.47 kV feeder), thereby eliminating the low frequency (LF) transformer for interconnection with the grid [67].

This work proposes an improved method to model the XFCS charging demand profile by considering EVs' departure from their parking places and associated initial SoCs, and it models their arrival times and SoCs at the time when they arrive at the station for recharging. Different from the literature [2, 19, 29], the charging demand profile of the XFCS is determined by using CPCV-based extreme fast charging of EVs at the station. Due to the substantial differences in the EVs' departure times and driving patterns during working days and non-working days, the XFCS demand profiles are separately modeled for a typical weekday and weekend day.

This study divides EVs into three categories: category-1 ($EVC_1$) includes privately owned EVs used for traveling from home to offices/workplaces, category-2 ($EVC_2$) privately owned EVs belonging to unemployed/retired persons for their personal use, and category-3 ($EVC_3$) corporately owned EVs used for work related travel and other purposes; this work assumes that $EVC_3$ includes company-owned electric buses, delivery trucks, and other large vehicles with larger battery packs compared to those used by $EVC_1$ and $EVC_2$ vehicles. Departure times for the EVs in the aforementioned categories is collected from the online database National Household Travel Survey (NHTS) [68]. It is assumed that for weekends, the departure time for EVs in all categories follows $\mathcal{N}_1(\mu_1, \sigma_1^2)$ normal probability distribution with mean $\mu_1$ and variance $\sigma_1^2$. For weekdays, it is assumed that the departure time for vehicles in categories $EVC_2$ and $EVC_3$ also follows the same $\mathcal{N}_1(\mu_1, \sigma_1^2)$ distribution. Vehicles in $EVC_1$, follow $\mathcal{N}_2(\mu_2, \sigma_2^2)$ normal distribution, with $\mu_2$ and variance $\sigma_2^2$, when departing from their parking places [69]. Table 1 presents the departure time assumptions for the considered EV categories.

Table 1. Probabilistic distribution for the departure time of EVs in different categories

| EV categories | Probabilistic Distribution for EVs Departure Time | |
|---|---|---|
| | Weekends | Weekdays |
| Category-1 ($EVC_1$) | $\mathcal{N}_1(\mu_1, \sigma_1^2)$ | $\mathcal{N}_2(\mu_2, \sigma_2^2)$ |
| Category-II ($EVC_2$) | $\mathcal{N}_1(\mu_1, \sigma_1^2)$ | $\mathcal{N}_1(\mu_1, \sigma_1^2)$ |
| Category-III ($EVC_3$) | $\mathcal{N}_1(\mu_1, \sigma_1^2)$ | $\mathcal{N}_1(\mu_1, \sigma_1^2)$ |

Per NHTS survey data, the daily driven mileage for EVs is assumed to follow the exponential probability distribution, as given by (1), where the coefficient $\varpi$ is obtained by the maximum likelihood method and its value is 0.0296 [69]. The exponential distribution $h(q_i)$ and histogram of a vehicle's daily driven mileage is shown in Fig. 2.

$$h(q_i) = \varpi e^{-\varpi q_i}, \quad q_i \geq 0. \tag{1}$$

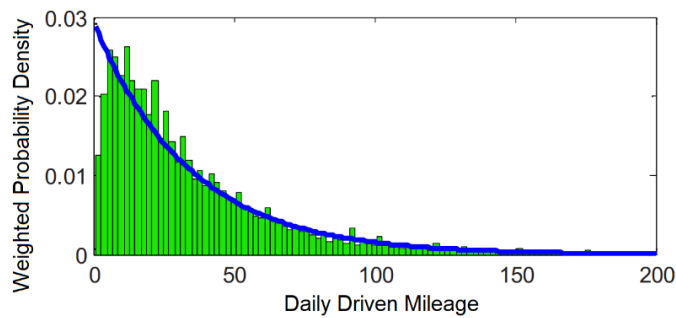

Fig. 2. Probability distribution and histogram of a vehicle's daily driven mileage [69]

In this work, the continuous probability density function is discretized, as presented in Fig. 2, and instead of using each possible mileage $q_i$ in the interval, the average driven mileage $Q_a$ (in miles) is computed for the interval $[Q_d, \ Q_{d+1}]$ using (2):

$$Q_a = \frac{Q_d + Q_{d+1}}{2}, \quad \forall Q_d = 1, 2, 3, \dots (miles) \tag{2}$$

where '$a$' is the index of EVs driven mileage, and $Q_d \ and \ Q_{d+1}$ are the distance values (in miles) used to obtain the average driven mileage ($Q_a$) in the interval $[Q_d, Q_{d+1}]$. Eq. (3) gives the probability $\mathcal{F}(Q_a)$ of driving the average distance of $Q_a$ in the interval $[Q_d, \ Q_{d+1}]$.

$$\mathcal{F}(Q_a) = \int_{Q_d}^{Q_{d+1}} \varpi e^{-\varpi q} \ dq. \tag{3}$$

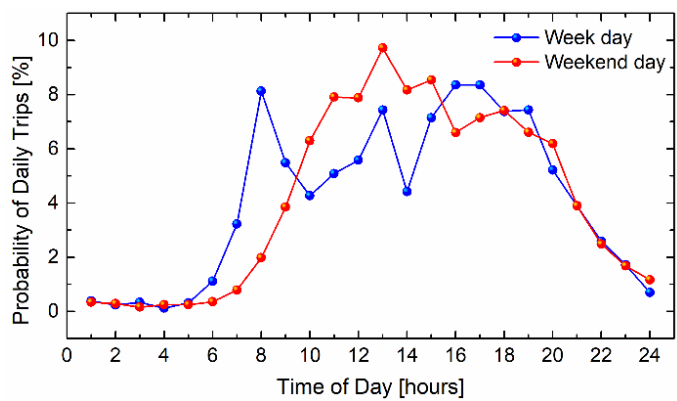

Fig. 3. Probability of daily trips for a typical weekday and weekend day

The XFCS demand modeling requires the probability of driving an EV at different times for both weekdays and weekends. Fig. 3 gives the hourly traffic distributions in the U.S. for a typical weekday and weekend day [2].

The arrival time and the arrival charge state for an EV can be obtained by using its departure time, state of charge at departure time, and the driving probability distribution at each hour around the clock, as given in Fig. 3. The state of charge of EV $i$ at time $t$ after driving a total distance of $Q_a$ miles is given by (4):

$$SoC_{i,a}^{EV}(t) = SoC_{i,a}^{EV}(t-1) - \frac{1}{\mathbb{C}_i^{EV}} \left( Q_a. \varphi^{EV}(t). E_i^{cpm} \right) \times 100. \tag{4}$$

This proposed model assumes a driver will immediately recharge the EV $i$ at the station when its SoC reaches the threshold $SOC_i^{thr}$. However, it should be noted that the $SOC_i^{thr}$ is not the same for each EV and follows a normal distribution $\mathcal{N}_3(\mu_3, \sigma_3^2)$ with mean $\mu_3$ and variance $\sigma_3^2$, thereby considering the heterogeneous actions of drivers in determining when and where to recharge their EVs when their SoCs reach the threshold [2]. The traveled distance for which an EV does not reach its $SoC_i^{thr}$, implying the EV will not go to the charging station for recharging, is represented by $\widetilde{Q_a}$, where $\widetilde{Q_a} \in Q_a - Q_a^*$. The weighted average of the arrival time ($\hat{t}_i^{arr}$) and state of charge ($\widehat{SoC}_i^{arr}$) of EVs arriving at the station after traveling $\mathcal{Q}_a^*$ distances, where $\mathcal{Q}_a^* \in \mathcal{Q}_a - \widetilde{\mathcal{Q}_a}$, is given by (5) and (6), respectively. The $\hat{t}_i^{arr}$ and $\widehat{SoC}_i^{arr}$ are computed using only those values of the traveled distance for which $\mathcal{Q}_a \leq \mathcal{Q}_a^*$. For the other values of $\mathcal{Q}_a$ (i.e., $\mathcal{Q}_a \leq \widetilde{\mathcal{Q}_a}$), EVs will not need to go to the charging station since their SoCs didn't reach the specified threshold; thus, $\mathcal{Q}_a \leq \widetilde{\mathcal{Q}_a}$ are not used for the computation of $\hat{t}_i^{arr}$ and $\widehat{SoC}_i^{arr}$ in (5) and (6).

$$\hat{t}_i^{arr} = \frac{\sum_{\mathcal{Q}_a}^{\mathcal{Q}_a^*}\left(\mathcal{F}(\mathcal{Q}_a^*) \cdot t_{i,\mathcal{Q}_a}^{arr}\right)}{\sum_{\mathcal{Q}_a}^{\mathcal{Q}_a^*}\left(\mathcal{F}(\mathcal{Q}_a^*)\right)} \tag{5}$$

$$\widehat{SoC}_i^{arr} = \frac{\sum_{\mathcal{Q}_a}^{\mathcal{Q}_a^*}\left(\mathcal{F}(\mathcal{Q}_a^*) \cdot SoC_{i,\mathcal{Q}_a}^{thr}\right)}{\sum_{\mathcal{Q}_a}^{\mathcal{Q}_a^*}\left(\mathcal{F}(\mathcal{Q}_a^*)\right)}. \tag{6}$$

To compute XFCS demand modeling, it is assumed that there are $r$ charging ports and $w$ waiting spots. Upon arrival at the station, $r$ EVs can be recharged simultaneously, and newly arrived EVs can stay in the waiting zone until a charging port becomes vacant. Once a charging port is available, EVs in the waiting zone can be recharged on a first-come-first-serve basis. If all the waiting spots are filled, new coming EVs may leave the station and find another station for recharging. The accepted EVs for recharging are denoted with $i^*$ and associated arrival time and state of charge with $\hat{t}_{i^*}^{arr}$ and $\widehat{SoC}_{i^*}^{arr}$, respectively. Depending on EV $i^*$ wait time before a charging port becomes available, the start time $t_{i^*}^{st}$ for recharging of that EV can be equal to or greater than its arrival time at the station.

The desired target state of charge for EV $i^*$, which started charging at $t_{i^*}^{st}$ for duration $t_{i^*}^{ch}$, is represented by (7). Note that EV users' heterogeneous behaviors in terms of terminating the charging process are considered by assuming that the $SoC_{i^*}^{target}$, for each $i^*$ EV, follows a normal probability distribution $\mathcal{N}_4(\mu_4, \sigma_4^2)$.

$$SoC_{i^*}^{target} = \widehat{SoC}_{i^*}^{arr} + \frac{E_{i^*}^{EV}(t)}{\mathfrak{E}_i^{EV}} \times 100. \tag{7}$$

The charging energy received by EV $i^*$ is given by (8). In this work, the CPCV charging method is utilized for extreme fast charging of EVs at the station. In the CPCV charging protocol, the EV battery is charged with a constant power in the CP mode until it reaches the cut-off voltage, after which the mode switches to CV mode wherein the voltage is held constant and charging power decreases [70]. The CPCV charging protocol is reported to have lower usable energy loss, higher charging efficiency, and lower cycle life aging of batteries especially in the case of fast charging (i.e., charging power > 50kW) [70-72]. Fig. 4 displays a sample CPCV profile (battery pack's recharge power, current, C-rate, and SoC) that was obtained by simulating the extreme fast charging of a 160-kWh battery pack. A C-rate is defined as the rate at which battery storage is charged/discharged with respect to its maximum capacity (C-rate unit is $h^{-1}$) [73].

$$E_{i^*}^{EV}(t) = \begin{cases} P^{XFCS} \times \Delta t, & CP \ mode \\ P(t) \times \Delta t, & CV \ mode \\ 0, & at \ other \ times. \end{cases} \tag{8}$$

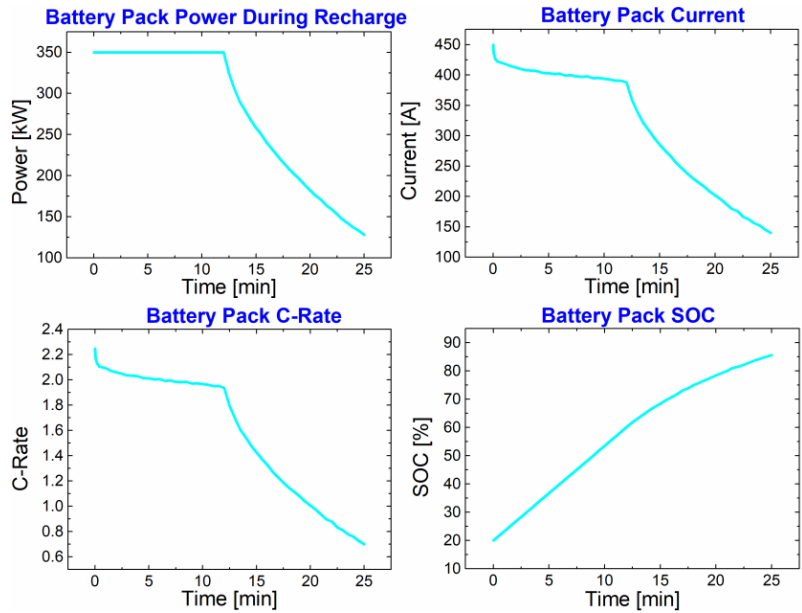

Fig. 4. Sample CPCV charging profile for 160-kWh battery pack with starting SoC=20% and desired SoC =85.6%.

The total energy demand profile of the XFCS is the aggregated charging energies of all the EVs served at the station on a typical weekday/weekend day and is given by (9):

$$E_{WD}(t) \ or \ E_{WND}(t) = \sum_{i^*}^{\mathcal{I}^*} E_{i^*}^{EV}(t). \tag{9}$$

## 3 Proposed Deterministic MILP Model

This section introduces mathematical formulations for the proposed deterministic MILP model to optimally size the XFCS components and obtain optimal energy station management. This paper investigates an XFCS that comprises of an active front end (AFE: three-phase AC/DC converter), DC/DC converters, a PV generation system, a Li-ion BESS, and three charging ports. The charging ports, PV system, and the BESS are tied to a common DC voltage bus [67], as shown in Fig. 1.

This work considers seasonal variations in the solar generation and electricity prices by representing a typical year with four seasons. Since the XFCS daily demand profile is computed for both weekdays and weekends, eight scenarios denoted by $j$ are used to characterize a typical year. The first four scenarios (i.e., $j = 1, 2, 3, 4$) represent the weekends, and scenarios five through eight (i.e., $j = 5, 6, 7, 8$) represent the weekdays during winter, spring, summer, and fall, respectively. The energy balance equation for the XFCS for scenario $j$ at time $t$ is given by (10). The efficiency losses in the AFE and DC/DC power converters are also considered, as observed in (10). The stored energy content in BESS for scenario $j$ at time $t$ is defined by (11) using its charge/discharge energies and the associated efficiencies [74]. The energy exchanged with the PDN for scenario $j$ at time $t$ is given by (12) in terms of energy imports/exports: the energy imported and exported from/to the PDN is respectively represented by positive and negative values. Simultaneous energy import and export is technically impossible and is enforced by (13) [75]. The Big-M method [44, 75] is utilized in (13b) and (13c) to linearize the bi-linear term appearing in (13a). The solar power generation is modeled in this work using the approach presented in [46, 76, 77]. Daily PV generation profiles for the considered scenarios at a time step duration of 1-min were calculated using the global irradiance and the daytime temperature profile for the selected location at Oak Ridge, Tennessee, USA. Eqs. (14-16) are used to model the PV system power profile ($P_{pv,j}(t)$) for scenario '$j$' at time '$t$', where (14) relates the $P_{pv,j}(t)$ with the per unit PV generation estimate ($\widehat{PV_{G,j}}(t)$)—which is modeled using (15) and (16) [76, 77]—and its rated capacity $P_{PV}$ [46]. Moreover, this paper assumes that full sunlight exposure is possible by installing the solar modules at open locations where shading from buildings, trees, and other objects does not reduce the PV system power generation [78]. The input parameters and solar irradiance profiles for the considered scenarios are presented in Section 5.2.

$$E_{g,j}^+(t) \times (\eta_{AC-DC} \times \eta_{DC-DC}) - \frac{E_{g,j}^-(t)}{(\eta_{AC-DC} \times \eta_{DC-DC})} = \widehat{E_{XFCS,j}}(t) + E_{ch,j}(t) - E_{dch,j}(t) - P_{pv,j}(t) \times \Delta t, \ \forall t \in \mathcal{T}, \forall j \in \mathcal{J} \tag{10}$$

$$E_{BESS,j}(t) = E_{BESS,j}(t-1) + E_{ch,j}(t) \times \eta_{ch} - \frac{E_{dch,j}(t)}{\eta_{dch}}, \ \forall t \in \mathcal{T}, \forall j \in \mathcal{J} \tag{11}$$

$$E_{g,j}^{AC}(t) = E_{g,j}^+(t) - E_{g,j}^-(t), \forall t \in \mathcal{T}, \forall j \in \mathcal{J} \tag{12}$$

$$E_{g,j}^+(t) \times E_{g,j}^-(t) = 0, \forall t \in \mathcal{T}, \forall j \in \mathcal{J} \tag{13a}$$

$$E_{g,j}^+(t) \leq \mathbb{M} \times u_{1,j}(t), \ \forall t \in \mathcal{T}, \forall j \in \mathcal{J} \tag{13b}$$

$$E_{g,j}^-(t) \leq \mathbb{M} \times \left(1 - u_{1,j}(t)\right), \forall t \in \mathcal{T}, \forall j \in \mathcal{J} \tag{13c}$$

$$P_{pv,j}(t) \leq P_{PV} \times \widehat{PV_{G,j}}(t), \ \forall t \in \mathcal{T}, \forall j \in \mathcal{J} \tag{14}$$

$$\widehat{PV_{G,j}}(t) = 0.92 \times IR_{G,j}(t) \times \frac{\left(1 - \beta \times \Delta T_j(t)\right)}{1000}, \forall t \in \mathcal{T}, \forall j \in \mathcal{J} \tag{15}$$

$$\Delta T_j(t) = \left|25 - \left(T_{Amb,j}(t) + (NOCT - 20) \times \frac{IR_{G,j}(t)}{800}\right)\right|, \forall t \in \mathcal{T}, \forall j \in \mathcal{J} \tag{16}$$

$$0 \leq E_{ch,j}(t) \leq P_{BESS}^{rated} \times \Delta t, \ \forall t \in \mathcal{T}, \forall j \in \mathcal{J} \tag{17}$$

$$0 \leq E_{dch,j}(t) \leq P_{BESS}^{rated} \times \Delta t, \ \forall t \in \mathcal{T}, \forall j \in \mathcal{J} \tag{18}$$

$$E_{ch,j}(t) \times E_{dch,j}(t) = 0, \forall t \in \mathcal{T}, \forall j \in \mathcal{J} \tag{19a}$$

$$E_{ch,j}(t) \leq \mathbb{M} \times u_{2,j}(t), \forall t \in \mathcal{T}, \forall j \in \mathcal{J} \tag{19b}$$

$$E_{dch,j}(t) \leq \mathbb{M} \times (1 - u_{2,j}(t)), \forall t \in \mathcal{T}, \forall j \in \mathcal{J} \tag{19c}$$

$$0 \leq (P_{BESS}^{rated} \times \Delta t) \leq E_{XFCS}^{max} \tag{20}$$

$$\underline{\gamma} \times P_{BESS}^{rated} \leq C_{BESS} \leq \overline{\gamma} \times P_{BESS}^{rated} \tag{21}$$

$$\sum_{t\in\mathcal{T}}\frac{E_{dch,j}(t)}{\eta_{dch}} = \sum_{t\in\mathcal{T}} E_{ch,j}(t)\times\eta_{ch}, \ \forall t\in\mathcal{T}, \forall j\in\mathcal{J} \tag{22}$$

$$E_{ch,j}(t)\leq (C_{BESS}-E_{BESS,j}(t)), \ \forall t\in\mathcal{T}, \forall j\in\mathcal{J} \tag{23}$$

$$E_{dch,j}(t)\leq E_{BESS,j}(t), \forall t\in\mathcal{T}, \forall j\in\mathcal{J} \tag{24}$$

$$-\pi \leq E_{BESS,j}(t)-E_{BESS,j}(t-1)\leq\pi, \forall t\in\mathcal{T}, \forall j\in\mathcal{J}. \tag{25}$$

$$SoC_{BESS,j}(t)=\frac{E_{BESS,j}(t)}{C_{BESS}}\times 100, \forall t\in\mathcal{T}, \forall j\in\mathcal{J} \tag{26}$$

The charging and discharging energies from the BESS are limited by kW sizing, as denoted by (17) and (18) [2, 79]. Moreover, simultaneous charging and discharging of the BESS is prohibited and given by (19). The big-M method is leveraged in (19b) and (19c) to linearize the bi-linear term appearing in (19a) [44]. The constraint in (20) limits the rated power of the BESS to be less than the maximum charging demand at the station. The practical limitations of the BESS, such as mutual dependence of its energy and power ratings, are represented by (21). The BESS stored energy is constrained to be equal for the start and end time of the optimization horizon, as given by (22). The charging and discharging energies of the BESS are constrained by available energy capacity and the BESS stored energy for scenario '$j$' at time '$t$', as given by (23) and (24), respectively. The ramp-down and ramp-up constraints are imposed on the charge/discharge energy of the BESS using (25) [80]. Eq. (26) gives the BESS state of charge in the percentage of its total energy capacity for scenario '$j$' at time '$t$'. Note that (26) makes the model non-linear, therefore it is used only to compute the $SoC_{BESS,j}(t)$ for presenting results after the optimization model is solved.

$$P_{g,avg,j}^{+}(k)=\left(\frac{1}{\xi}\right)\sum_{t=1+\xi\times(k-1)}^{k\times\xi}\left(\frac{E_{g,j}^{+}(t)}{\Delta t}\right) \ \ \forall k\in\mathcal{K}, \forall j\in\mathcal{J} \tag{27}$$

$$P_{g,daily,j}^{max}=max\left(P_{g,avg,j}^{+}(k)\right), \forall k\in\mathcal{K}, \forall j\in\mathcal{J} \tag{28}$$

$$P_{g,mo,s}^{max}=max\left(P_{g,daily,s}^{max},P_{g,daily,s+\hat{s}}^{max}\right), \forall s\in\mathcal{S} \tag{29}$$

$$P_{g,an}^{max}=max\left(P_{g,mo,s}^{max}\right), \forall j\in\mathcal{J} \tag{30}$$

$$P_{g,daily,j}^{max}\geq P_{g,avg,j}^{+}(k), \forall k\in\mathcal{K}, \forall j\in\mathcal{J} \tag{31a}$$

$$\begin{cases} P_{g,mo,s}^{max}\geq P_{g,daily,s}^{max} \\ P_{g,mo,s}^{max}\geq P_{g,daily,s+\hat{s}}^{max} \end{cases}, \forall s\in\mathcal{S} \tag{31b}$$

$$P_{g,an}^{max}\geq P_{g,mo,s}^{max}, \forall s\in\mathcal{S}. \tag{31c}$$

To compute the monthly and annual demand charges, the daily average power imported from the PDN (averaged over $\xi$-mins window) is computed using (27), and its maximum value for scenario $j$ during 24-hours is obtained using (28). Since each season is characterized by one weekday and one weekend day, the maximum monthly average power imported from the PDN for season $s$, is computed using (29). Finally, the maximum annual average power import is given by (30). In (29)-(30), the '$max$' function is nonlinear; therefore, (31) is used to replace it in the proposed linear formulations. Note the addition of (31) is sufficient to replace the '$max$' function in the proposed model because the nature of the objective function (40) guarantees that the smallest feasible values of $P_{g,daily,j}^{max}$, $P_{g,mo,s}^{max}$, and $P_{g,an}^{max}$ will always be selected:

$$\Psi = \frac{(\mathbb{D}/_S)}{c_{BESS}} \cdot \sum_{t \in \mathcal{T}} \left[ \frac{2}{7} \sum_{j=1}^{4} \left( \frac{E_{dch,j}(t)}{\eta_{dch}} \right) + \frac{5}{7} \sum_{j=5}^{8} \left( \frac{E_{dch,j}(t)}{\eta_{dch}} \right) \right] = \frac{(\mathbb{D}/_S)}{c_{BESS}} \cdot \sum_{t \in \mathcal{T}} \left[ \begin{array}{c} \frac{2}{7} \sum_{j=1}^{4} \left( \eta_{ch} \times E_{ch,j}(t) \right) + \\ \frac{5}{7} \sum_{j=5}^{8} \left( \eta_{ch} \times E_{ch,j}(t) \right) \end{array} \right] \quad \forall t \in \mathcal{T}, \forall j \in \mathcal{J}. \quad (32)$$

Studies [81-83] have been proposed in the literature that model the battery degradation accurately in a very detailed manner. However, because the proposed models are non-linear and involve multiple variables, they are computationally burdensome for planning studies which is not acceptable. Introducing non-linear formulations in the optimal sizing model can also lead to a local optimum [84, 85]. Furthermore, to simulate the accurate capacity fading phenomenon, the planning model would need to be solved for the entire planning horizon which is extremely computation-expensive, especially when the simulation time step uses minute-scale time granularity as used in this work. Therefore, this paper utilizes a different BESS degradation method, adopted from [2, 65, 66, 84] and is suitable for planning studies, that considers the cycle-life degradation characteristics of the Li-ion based battery energy storage system (BESS) in terms of its depth of discharge (the DoD at which it will be operated during the project lifetime) and corresponding maximum allowable cycles using the relationship presented in Fig. 5.

The DoD per cycle and charge/discharge cycles of a BESS are the chief operating factors contributing to its cycle-life degradation during its calendar lifetime. BESS cycle-life degradation is a key aspect to consider for obtaining the optimal DoD and the maximum allowed number of cycles to make sure that the BESS will not be replaced during the considered project lifetime. Therefore, in addition to BESS investment costs ($/kW and $/kWh), it is also important to take its number of charge/discharge cycles and DoD per cycle into account in the BESS planning studies [2, 65, 66]. One complete cycle of the BESS consists of one full charge cycle and a discharge cycle. In this work, the number of BESS charge and discharge cycles in a day are equal because the BESS stored energy is constrained to be equal for start and end times in the day. In the existing literature on BESS sizing [2, 65, 66], one discharge cycle is completed when the BESS energy capacity is discharged from 100% to 0% continuously all from one discharge (i.e., no charging can happen before finishing a discharge cycle). On the contrary, in this study, one discharge cycle is completed when an equivalent amount of energy that amounts to 100% of the ESS capacity is discharged but not essentially all from one discharge. Thus, this paper offers a more practical approach for tallying BESS cycles, as given by (32). In (32), the annual cycle count is obtained using cumulative charging/discharging energy.

This work uses a piece-wise linear approximation of the Li-ion BESS cycle life curve shown in Fig. 5, which was taken from [86] and approximated using expressions in (33). It should be noted that in (33), a maximum of two adjacent non-zero $\rho$'s are in the final solution, which can be realized in AIMMS (an optimization modeling tool) by specifying a special order set of type-2 (SOS2) in the property attribute of constraints given in (33) [87, 88]. The total number of BESS charge/discharge cycles is constrained to be less than the allowed cycles corresponding to the optimal DoD, as given by (34). The constraint in (35) sets the lower limit of BESS stored energy based on the determined optimal DoD.

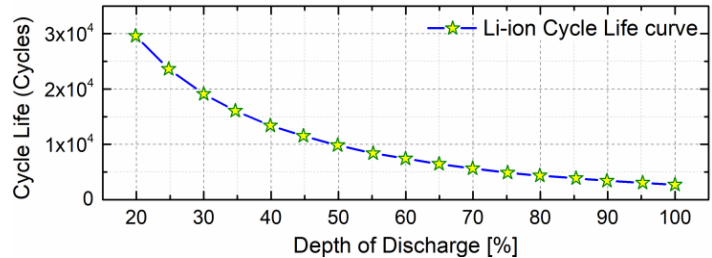

Fig. 5. Cycle life curve of Li-ion based BESS

$$\begin{cases} \mathfrak{I}^{a}(DoD) = \sum_{s\in\mathbb{S}}\big(\rho(s)\times\mathfrak{I}(DoD(s))\big) \\ DoD = \sum_{s\in\mathbb{S}}\big(\rho(s)\times DoD(s)\big) \\ \sum_{s\in\mathbb{S}}\big(\rho(s)\big) = 1 \end{cases} \tag{33}$$

$$\Psi \le \mathfrak{I}^{a}(DoD) \tag{34}$$

$$(1-DoD)\times C_{BESS} \le E_{BESS,j}(t) \le C_{BESS}. \tag{35}$$

The constraints in (32) and (35) contain non-linear terms. Rearranging (32) and (35) yields bi-linear terms, i.e., '$\Psi \times C_{BESS}$' and '$DoD \times C_{BESS}$' respectively given by (36) and (37), which can be linearized using McCormick relaxations. A McCormick envelope can form the tightest convex hull of bi-linear functions [89]. Interested readers should refer to [90-92] for detailed discussions about standard McCormick relaxations and tightening piece-wise McCormick relaxations for bi-linear terms.

$$\Psi \times C_{BESS} = \big(\mathcal{D}/\mathcal{S}\big)\sum_{t\in\mathcal{T}}\left[\frac{2}{7}\sum_{j=1}^{4}\left(\frac{E_{dch,j}(t)}{\eta_{dch}}\right) + \frac{5}{7}\sum_{j=5}^{8}\left(\frac{E_{dch,j}(t)}{\eta_{dch}}\right)\right] = \big(\mathcal{D}/\mathcal{S}\big)\sum_{t\in\mathcal{T}}\begin{bmatrix}\frac{2}{7}\sum_{j=1}^{4}\big(\eta_{ch}\times E_{ch,j}(t)\big) + \\ \frac{5}{7}\sum_{j=5}^{8}\big(\eta_{ch}\times E_{ch,j}(t)\big)\end{bmatrix}, \forall t\in\mathcal{T}, \forall j\in\mathcal{J} \tag{36}$$

$$(C_{BESS} - DoD\times C_{BESS}) \le E_{BESS,j}(t) \le C_{BESS} \tag{37}$$

Let $\Upsilon = \Psi \times C_{BESS}$

$$\begin{cases} \Upsilon \geq \Psi.\underline{C_{BESS}} + \underline{\Psi}.C_{BESS} - \underline{\Psi}.\underline{C_{BESS}} \\ \Upsilon \geq \Psi.\overline{C_{BESS}} + \overline{\Psi}.C_{BESS} - \overline{\Psi}.\overline{C_{BESS}} \\ \Upsilon \leq \Psi.\overline{C_{BESS}} + \underline{\Psi}.C_{BESS} - \underline{\Psi}.\overline{C_{BESS}} \\ \Upsilon \leq \Psi.\underline{C_{BESS}} + \overline{\Psi}.C_{BESS} - \overline{\Psi}.\underline{C_{BESS}} \end{cases} \tag{38}$$

Let $\omega = DoD \times C_{BESS}$

$$\begin{cases} \omega \geq DoD.\underline{C_{BESS}} + \underline{DoD}.C_{BESS} - \underline{DoD}.\underline{C_{BESS}} \\ \omega \geq DoD.\overline{C_{BESS}} + \underline{DoD}.C_{BESS} - \underline{DoD}.\overline{C_{BESS}} \\ \omega \leq DoD.\overline{C_{BESS}} + \underline{DoD}.C_{BESS} - \underline{DoD}.\overline{C_{BESS}} \\ \omega \leq DoD.\underline{C_{BESS}} + \overline{DoD}.C_{BESS} - \overline{DoD}.\underline{C_{BESS}} \end{cases}. \tag{39}$$

Consequently, the linearized expressions (38) and (39) replace the bilinear terms of (36) and (37), resulting in a standard MILP model. Using the approach presented in [92], MILP formulations are used to obtain the tightest piece-wise McCormick relaxation corresponding to (38), as given in Appendix A; formulations can also be derived for (39) using the same approach. In (38) and (39), $C_{BESS} \in [\underline{C_{BESS}}, \overline{C_{BESS}}]$, $\Psi \in [\underline{\Psi}, \overline{\Psi}]$, and $DoD \in [\underline{DoD}, \overline{DoD}]$.

The objective function (OF) of the proposed formulations is to minimize the total yearly XFCS cost that consists of three main components: monthly and yearly demand charges, investment and O&M costs for the PV system and the BESS, and the operating cost for the station, as expressed in (40). The decision vector of the proposed formulations is $\mathcal{V} = [C_{BESS}, P_{BESS}^{rated}, DoD, \psi, E_{BESS,j}(t), P_{PV}, P_{pv,j}(t), E_{g,j}^{AC}(t), P_{g,mo,s}^{max}, P_{g,an}^{max}]$.

The annualized investment costs of the BESS and the PV system are obtained using $CF_{ann}$ from (41) [2]. The complete DO optimization model consists of the OF (40) and the constraints (10-39).

$$\min_{\mathcal{V}}. \left( \overbrace{SF^M \sum_{s \in \mathcal{S}} \left( P_{g,mo,s}^{max} \times \lambda_{MDC} \right)}^{\text{Monthly demand charges (MDC)}} + \overbrace{\left( P_{g,an}^{max} \times \lambda_{ADC} \right)}^{\text{Annual demand charges (ADC)}} + \overbrace{\left( C_{BESS}(CE_{BESS}^{cap} + CE_{BESS}^{inst}) + P_{BESS}^{rated} \left( CP_{BESS}^{cap} + C_{BESS}^{O\&M} \right) \right)}^{\text{Investment cost, installation cost, and O\&M cost of BESS (IC\&OM}^{BESS})} \right.$$

$$\left. + \overbrace{\left( P_{PV}(CP_{PV}^{cap} + C_{PV}^{O\&M}) \right)}^{\text{Investment cost and O\&M cost of PV system (IC\&OM}^{PV})} + \overbrace{\left( (\mathcal{D}/\mathcal{S}) \sum_{t \in \mathcal{T}} \begin{pmatrix} \frac{2}{7}\sum_{j=1}^{4} \left( E_{g,j}^{AC}(t) \times \hat{\lambda}_{E,j}(t) \right) + \\ \frac{5}{7}\sum_{j=5}^{8} \left( E_{g,j}^{AC}(t) \times \hat{\lambda}_{E,j}(t) \right) \end{pmatrix} \right)}^{\text{Cost of energy exchanged with PDN (OpC)}} \right),$$

$$\forall t \in \mathcal{T}, \forall j \in \mathcal{J} \tag{40}$$

$$CF_{ann} = \frac{z \times (1+z)^{\mathcal{L}}}{(1+z)^{\mathcal{L}} - 1}. \tag{41}$$

## 4 Robust Optimization based MILP Model

The RO approach is an alternative method for considering uncertainties in planning problems without needing accurate information about uncertain data probability distribution functions (PDFs). Instead, it only requires limited information about uncertain data in the form of uncertainty confidence bounds. The RO solution is workable for all uncertainty realizations within the specified bounds, and it is obtained at the worst-case realization of uncertain parameters [45, 48].

### 4.1 Uncertainty Modeling of Electricity Market Price

This section outlines the RO modeling to account for uncertainties in wholesale electricity prices. The DO model (40) can be recast as a min-max-min RO model, given by (42). The max-min structure of (42) relates to the fact that, while the inner term minimizes the OF over decision variables, the outer term realizes the uncertainty such that it occasions the worst possible impact on the OF. The decision vector is $\mathcal{V} = [C_{BESS}, P_{BESS}^{rated}, DoD, \psi, E_{BESS,j}(t), P_{PV}, P_{pv,j}(t), E_{g,j}^{AC}(t), P_{g,mo,s}^{max}, P_{g,an}^{max}]$.

$$\min_{\mathcal{V}} \left( MDC + ADC + IC\&OM^{BESS} + IC\&OM^{PV} \right) + \max_{\lambda_{E,j}^{\ell}(t) \in U^{\lambda}} \min_{\mathcal{V}} \left( \binom{\mathcal{D}}{S} \sum_{t \in \mathcal{T}} \begin{pmatrix} \frac{2}{7} \sum_{j=1}^{4} \left( E_{g,j}^{AC}(t) \times \lambda_{E,j}^{\ell}(t) \right) + \\ \frac{5}{7} \sum_{j=5}^{8} \left( E_{g,j}^{AC}(t) \times \lambda_{E,j}^{\ell}(t) \right) \end{pmatrix} \right),$$

$$\forall t \in \mathcal{T}, \forall j \in \mathcal{J} \tag{42}$$

$$U^{\lambda} = \left\{ \begin{matrix} \lambda_{E,j}^{\ell}(t) \in R^{+} : \underline{\Gamma^{\lambda}} \leq \frac{\lambda_{E,j}^{\ell}(t)}{\hat{\lambda}_{E,j}(t)} \leq \overline{\Gamma^{\lambda}} \\ \lambda_{E,j}^{\ell}(t) \in \left[ \underline{\lambda_{E,j}}(t), \overline{\lambda_{E,j}}(t) \right] \\ \overline{\lambda_{E,j}}(t) = (1 + \Phi) \times \hat{\lambda}_{E,j}(t) \\ \underline{\lambda_{E,j}}(t) = (1 - \Phi) \times \hat{\lambda}_{E,j}(t) \end{matrix} \right\}, \forall t \in \mathcal{T}, \forall j \in \mathcal{J}. \tag{43}$$

Using the approach presented in [93], the min-max-min RO structure is reformulated into a standard minimization problem, as given by (44), using duality properties and linearization of the constraints. Interested readers should see [93] for a detailed discussion about the implemented approach. This approach transforms the min-max-min structure-based model into a tractable minimization problem and provides a way to adjust the degree of risk-aversion in the final solution by tuning the robust parameter. The bounds of the electricity price are given by (43). Furthermore, the lower and upper limits on the electricity price signal are enforced using the uncertainty budget, denoted by $\underline{\Gamma^{\lambda}}$ and $\overline{\Gamma^{\lambda}}$ for managing the degree of conservatism for $U^{\lambda}$. The uncertainty budget '$\Gamma^{\lambda}$' can take any value between $[0, \mathcal{T}]$, where $\mathcal{T}$ is total minutes in a day. Increasing the $\Gamma^{\lambda}$ makes the robust solution more conservative, with $\Gamma^{\lambda} = \mathcal{T}$ the solution is most conservative, while $\Gamma^{\lambda} = 0$ yields the least conservative solution, i.e., deterministic case. Put simply, $\Gamma^\lambda$ controls the trade-off between economic performance and the robustness against uncertainties in the final solution:

$$min. \left(MDC + ADC + IC\&OM^{BESS} + IC\&OM^{PV}\right) + \left(\left(\tfrac{1}{S}\right)\sum_{t\in\mathcal{T}}\left(\begin{array}{l}\frac{2}{7}\sum_{j=1}^{4}\left(E_{g,j}^{AC}(t)\times\hat{\lambda}_{E,j}(t)\right)+\\\frac{5}{7}\sum_{j=5}^{8}\left(E_{g,j}^{AC}(t)\times\hat{\lambda}_{E,j}(t)\right)\end{array}\right)\right) + \left(\Gamma^\lambda\times\alpha^{\lambda_E}+\sum_{t\in\mathcal{T}}\beta^{\lambda_E}(t)\right),$$

$$\forall t\in\mathcal{T}, \forall j\in\mathcal{J} \quad (44)$$

s.t. the constraints (10-39), (43), and (45).

$$\alpha^{\lambda_E}+\beta^{\lambda_E}(t) \geq \left(\tfrac{1}{S}\right)\left(\frac{2}{7}\sum_{j=1}^{4}\left(\frac{\overline{\lambda_{E,j}}(t)-\underline{\lambda_{E,j}}(t)}{2}\times\varsigma_j^{\lambda_E}(t)\right)+\frac{5}{7}\sum_{j=5}^{8}\left(\frac{\overline{\lambda_{E,j}}(t)-\underline{\lambda_{E,j}}(t)}{2}\times\varsigma_j^{\lambda_E}(t)\right)\right),$$

$$\alpha^{\lambda_E}\geq 0,$$

$$\beta^{\lambda_E}(t)\geq 0,$$

$$\varsigma_j^{\lambda_E}(t)\geq 0,$$

$$\varsigma_j^{\lambda_E}(t)\geq E_{g,j}^{AC}(t), \quad \forall t\in\mathcal{T}, \forall j\in\mathcal{J}. \quad (45)$$

where $\alpha^{\lambda_E}$ and $\beta^{\lambda_E}(t)$ are dual variables of the original DO model, and $\varsigma_j^{\lambda_E}(t)$ is an auxiliary variable that is used to achieve the corresponding linear formulations.

### 4.2 Uncertainty Modeling of XFCS Demand and PV System Generation

In this section, uncertainty modeling of the XFCS demand profile and the PV system power generation is addressed. Similar to Section 4.1, the DO model (40) is recast as a min-max-min RO structure, as given by (46). The uncertainty bounds for both input parameters are given by (47) and (48). In addition, the uncertainty budgets for charging demand and PV generation, respectively, are denoted by $\Gamma^D$ and $\Gamma^{PV}$, manage the degree of conservatism in the final solution. Note that the optimal rating of the PV system is one of the decision variables and the uncertainty exists in the per unit PV generation data, which is modeled using (13) and (14). The decision vector is $\mathcal{V}=[C_{BESS}, P_{BESS}^{rated}, DoD, \psi, E_{BESS,j}(t), P_{PV}, P_{pv,j}(t), E_{g,j}^{AC}(t), P_{g,mo,s}^{max}, P_{g,an}^{max}]$.

$$min. \left(MDC + ADC + IC\&OM^{BESS} + IC\&OM^{PV}\right) + \max_{\substack{E_{XFCS,j}^{d}(t)\in U^D, PV_{G,j}^p(t)\in U^{PV}}} min. \left(\left(\tfrac{1}{S}\right)\sum_{t\in\mathcal{T}}\left(\begin{array}{l}\frac{2}{7}\sum_{j=1}^{4}\left(E_{g,j}^{AC}(t)\times\hat{\lambda}_{E,j}(t)\right)+\\\frac{5}{7}\sum_{j=5}^{8}\left(E_{g,j}^{AC}(t)\times\hat{\lambda}_{E,j}(t)\right)\end{array}\right)\right),$$

$$\forall t \in \mathcal{T}, \forall j \in \mathcal{J} \quad (46)$$

$$U^D = \left\{ \begin{array}{l} E^d_{XFCS,j}(t) \in R^+ : \underline{\Gamma^D} \leq \dfrac{E^d_{XFCS,j}(t)}{\widehat{E_{XFCS}}(t)} \leq \overline{\Gamma^D} \\[2mm] E^d_{XFCS,j}(t) \in \left[ \underline{E_{XFCS,j}}(t), \overline{E_{XFCS,j}}(t) \right] \\[2mm] \overline{E_{XFCS,j}}(t) = (1 + \varrho) \times \widehat{E_{XFCS,j}}(t) \\[2mm] \underline{E_{XFCS,j}}(t) = (1 - \varrho) \times \widehat{E_{XFCS,j}}(t) \end{array} \right\}, \ \forall t \in \mathcal{T}, \forall j \in \mathcal{J} \quad (47)$$

$$U^{PV} = \left\{ \begin{array}{l} PV^p_{G,j}(t) \in R^+ : \underline{\Gamma^{PV}} \leq \dfrac{PV^p_{G,j}(t)}{\widehat{PV_{G,j}}(t)} \leq \overline{\Gamma^{PV}} \\[2mm] PV^p_{G,j}(t) \in \left[ \underline{PV_{G,j}}(t), \overline{PV_{G,j}}(t) \right] \\[2mm] \overline{PV_{G,j}}(t) = (1 + \varepsilon) \times \widehat{PV_{G,j}}(t) \\[2mm] \underline{PV_{G,j}}(t) = (1 - \varepsilon) \times \widehat{PV_{G,j}}(t) \end{array} \right\}, \forall t \in \mathcal{T}, \forall j \in \mathcal{J}. \quad (48)$$

In the RO model, using the approach from [93], the energy balance constraint of (10) is transformed into (49), with additional constraints of (50) and (51):

$$E^+_{g,j}(t) \times (\eta_{AC-DC} \times \eta_{DC-DC}) - \frac{E^-_{g,j}(t)}{(\eta_{AC-DC} \times \eta_{DC-DC})} = \left( \Gamma^D \times \alpha^D + \beta^D_j(t) + \widehat{E_{XFCS,j}}(t) + E_{ch,j}(t) - E_{dch,j}(t) - \left( -\Gamma^{PV} \times \alpha^{PV} - \right. \right.$$
$$\left. \left. \beta^{PV}_j(t) + P_{pv,j}(t) \right) \Delta t \right), \ \forall t \in \mathcal{T}, \forall j \in \mathcal{J} \quad (49)$$

$$\alpha^D + \beta^D_j(t) \geq \frac{\left( \overline{E_{XFCS,j}}(t) - \underline{E_{XFCS,j}}(t) \right)}{2} \times \varsigma^D_j(t),$$
$$\alpha^D \geq 0,$$
$$\beta^D_j(t) \geq 0,$$
$$\varsigma^D_j(t) \geq 0,$$
$$\varsigma^D_j(t) \geq 1 \quad \forall t \in \mathcal{T}, \forall j \in \mathcal{J} \quad (50)$$

$$\alpha^{PV} + \beta^{PV}_j(t) \geq \frac{\left( \overline{PV_{G,j}}(t) - \underline{PV_{G,j}}(t) \right)}{2} \times \varsigma^{PV}_j(t),$$
$$\alpha^{PV} \geq 0,$$
$$\beta^{PV}_j(t) \geq 0,$$
$$\varsigma^{PV}_j(t) \geq 0,$$
$$\varsigma^{PV}_j(t) \geq P_{PV}, \ \forall t \in \mathcal{T}, \forall j \in \mathcal{J}. \quad (51)$$

where $\alpha^D$, $\beta^D_j(t)$, $\alpha^{PV}$, and $\beta^{PV}_j(t)$ are the dual variables of the original DO model related to charging demand and solar generation uncertainties; $\varsigma^D_j(t)$ and $\varsigma^{PV}_j(t)$ are the auxiliary variables utilized for obtaining the corresponding linear expressions.

The complete RO-based MILP model comprises of the OF (44) and the constraints (11-39), (43), (45), and (47-51).

## 5  Case Study Simulations

### 5.1  Simulations of the XFCS Demand Model

In XFCS demand modeling, the total number of EVs is assumed to be 100 [2]. The following assumptions are made to calculate the XFCS demand: there is only one XFCS in the region under study. If an EV '$i$' originating from this region travels within the region and its SoC reaches the threshold $SOC_i^{thr}$, specified in the demand model, it will get recharged at the XFCS under study. If EV users travel from other regions and SoCs of their EVs reach pre-specified thresholds, they will also recharge EVs at this XFCS if they cannot reach back to their regions of origin without being recharged, otherwise they will recharge EVs at the charging stations of their own regions. Additionally, it is assumed that the number of EVs traveling out of the region under study are the same as those traveling in the region, such that, the total number of considered EVs in the region under study remains 100. Lastly, the XFCS demand modeling approach is not applicable and needs to be revised if there are multiple charging stations in the region.

The percentages of EVs belonging to $EVC_1$, $EVC_2$, and $EVC_3$ categories are taken as 61%, 30%, and 9%, respectively [94]. The normal probability distribution values of departure times for $EVC_2$/$EVC_3$, and $EVC_1$ are taken from [69] and are $\mathcal{N}_1(13:51, 5:12)$ and $\mathcal{N}_2(06:52, 1:18)$, respectively. The battery capacities of EVs in the $EVC_1$ and $EVC_2$ are assumed to be 100-kWh, with energy consumptions of 0.35 kWh/mile [95, 96], which is used in the Tesla Model S 100D (2018) and Tesla Model X 100D (2018) [97, 98]. The battery capacity of 160-kWh, with energy consumption of 2.0 kWh/mile [99], is assumed for EVs in the $EVC_3$ [100-102]. The normal probability distribution representing the threshold $SOC_i^{thr}$ is taken from [103] and is $\mathcal{N}_3(30,15)$. Moreover, to avoid deep-discharge and over-charging and to enhance battery life, the state of charge for all EVs is assumed to be constrained within the range of 10-90%, at all times [104, 105]. The $SoC_{i^*}^{target}$ also follows a normal distribution $\mathcal{N}_4(80,10)$ [106]. For the purpose of simulations, it is assumed that there are $r = 3$ charging ports and $w = 5$ waiting spots in the studied charging station. Obtaining the optimal number of charging and waiting spots is out of this study's scope. The kW rating of each charging port in the studied XFCS is taken as 350 kW [10, 107].

The obtained XFCS demand profiles for a typical weekday and weekend are displayed in Fig. 6. The XFCS demand has two peaks for a typical weekday: one peak is observed in the morning time and the second in the evening/nighttime. For the weekend, the majority of the EVs get charged in the afternoon and evening times. In this work, the XFCS charging demand is assumed to be the same for all seasons.

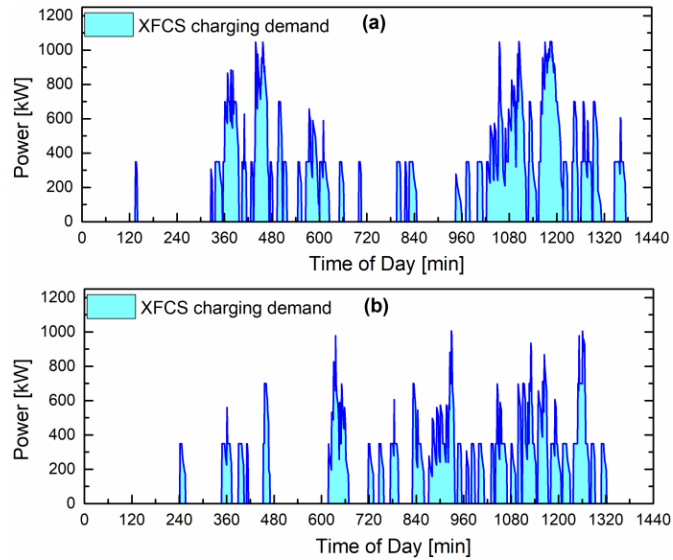

Fig. 6. XFCS daily charging demand profile for a typical (a) weekday and (b) weekend with $r$=3 charging ports and $w$=5 waiting spots.

### 5.2 Simulation of the XFCS Sizing Model

The proposed MILP sizing model was implemented in AIMMS 4.74.1.0 and solved using CPLEX 12.10 solver [88]. The model was implemented on a computer with 32 GB of RAM and a 3.20 GHz Intel® Core™ i7 processor. The solutions for the DO and RO models were obtained in ~26 and ~102 seconds, respectively.

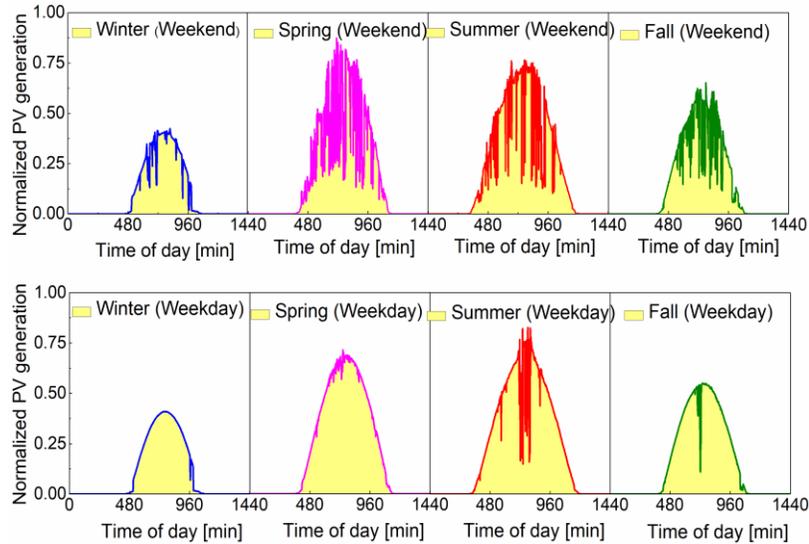

Fig. 7. Per unit estimate of PV generation profiles for each season

In this work, one year is characterized by eight scenarios to include the effects of seasonal differences in the PV generation profiles and electricity prices, as discussed in Section 3. To signify seasonal variations in the per unit estimate of PV power generation ($\widehat{PV_{G,j}}(t)$), the solar irradiance and ambient temperature data were collected for four seasons (for both weekends and weekdays) for the year 2019 and location Oak Ridge, Tennessee, USA (Latitude: 35.92996º North

Longitude: 84.30952° West) from an online database of the National Renewable Energy Laboratory (NREL) [108]. Fig. 7 illustrates the per unit estimate of the PV generation profiles for each season. This work assumes the total area of the XFCS is 2000 $m^2$ [44], therefore the PV generation system maximum power rating is restricted to 300 kW [109]. Moreover, the capacity of the PDN feeder and ratings of power electronic converters (PECs) are assumed to be enough to satisfy the XFCS peak power demand.

Among different pricing mechanisms, Time of Use (ToU), Real-time Pricing (RTP), and Critical Peak Pricing (CPP) are the most appropriate pricing methods in US energy markets [110]. RTP increases the BESS's revenue, but it generates the highest price variation for the prosumer, according to [111]. TOU and CPP, on the other hand, only bring a modest amount of profit to the retailer's bottom line. The optimal scheduling problem [74, 79, 112] and system planning [113, 114] of various players in power markets have been studied in the literature considering the availability of EV and charging stations. The superiority of various demand response (DR) programs is dependent on the numerous case studies and the examined problem, according to the results and explanations in the literature. To provide a more realistic case study, this paper employed the RTP pricing mechanism. Fig. 8 shows the electricity price curves for four seasons taken from [44].

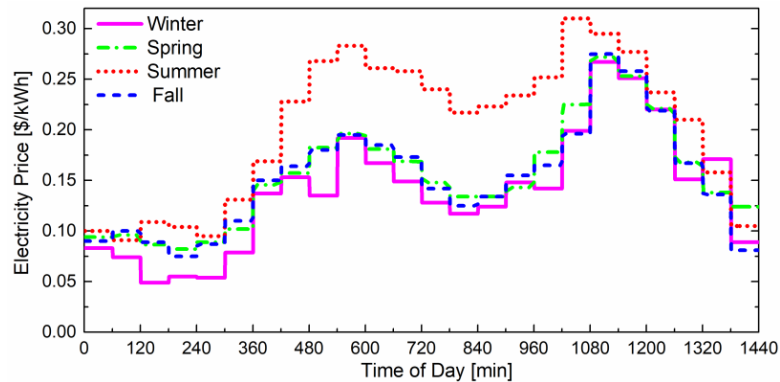

Fig. 8. Electricity market prices for four seasons

The input parameters related to BESS and PV generation system are taken from [2, 76] and are presented in Table 2.

Table 2. Simulation Parameters

| Parameter | Value | Parameter | Value |
|---|---|---|---|
| BESS technology | Li-ion | Mounting status of PV module and tilt angle | Free-standing and 34 degrees |
| BESS power rating cost [$/kW] | 300 | PV system losses [%] (estimated) | 14 |
| BESS annual O&M cost [$/kW] | - | PV NOCT [°C] | 45 |
| BESS energy rating cost [$/kWh] | 695 | $\eta_{AC-DC}, \eta_{DC-DC}$ [%] | 95 |
| BESS installation cost [$/kWh] | 3.6 | Optimization horizon [years] | 1 |
| BESS lifetime | 20 | $z$ [%] | 4 |
| [years] | | | |
| BESS $\gamma^{min}, \gamma^{max}$ | 1, 8 | $\xi$ [min] | 15 |
| BESS $\eta_{ch}, \eta_{dch}$ [%] | 98 | $\mathcal{K}$ | 96 |
| $\pi$ [kWh/min] | 20 | $\lambda_{ADC}, \lambda_{MDC}$ [$/kW] | 18, 10 |
| PV technology | c-Si | $\mathfrak{D}, \mathcal{T}, \mathcal{S}, \mathcal{J}$ | 365, 1440, 4, 8 |
| PV power rating cost [$/kW] | 2277 | $SF^M$ | 3 |
| PV system O&M cost [$/kW/year] | 21 | $\Delta t$ [h] | 1/60 |
| PV system lifetime [years] | 25 | $\mathcal{L}$ [years] | 20 |
| $\hat{\beta}$ | 0.007 | | |

*5.3 Results and Discussion*

*5.3.1 XFCS Sizing using Deterministic Optimization Model*

This section presents sizing of the XFCS components and optimal energy management of the XFCS using the DO model (without considering any uncertainty in the input data) for a planning horizon of 20 years. The investment costs of the BESS and PV system are obtained with the help of expression (41) using the interest rate and lifetime of the project. Table 3 presents comparisons of the sizing results and savings in the total annualized cost of the station for the following three cases:

- Base Case: Without considering the application of BESS and PV system in the station

- Case I: XFCS sizing with BESS and PV system: without considering the BESS life degradation

- Case II: XFCS sizing with BESS and PV system: by considering the BESS life degradation

In the Base Case, the total annualized cost of the station is computed as a baseline for appraising the savings obtained from Cases I and II. Between Cases I and II, it is found that the BESS energy and power sizing are higher for Case I; moreover, the optimal DoD and annual charge/discharge cycles are also higher in Case I. The increased BESS power and energy ratings in Case I can be attributed to the fact that with no limitations on the allowed annual BESS cycles (i.e., with no degradation considerations), the energy arbitrage can be performed to a higher degree and earn more revenue and annual savings for the station compared to Case II, which is evident in Table 3. Case I promises annual savings of 34.66% in contrast to 23.12% expected in Case II, so having a larger BESS is justified in Case I. The maximum monthly and annual average power imported from the PDN and the resulting demand charges are higher in Case I (approx. 35% higher) in comparison with Case II. The justification for this is that the revenue earned via energy arbitrage overshadows the reduction in the demand charges using BESS; therefore, in Case I, the BESS is utilized more often to exploit the energy arbitrage opportunities by charging during the low-price hours and discharging during the peak-price hours, thereby resulting in ~58% lower operating cost compared to Case II.

Table 3. Simulation results of XFCS sizing with a project lifetime of 20 years

| | Base Case (w/o PV & BESS) | Case I | Case II |
|---|---|---|---|
| $C_{BESS}$ [kWh] | - | 2207.47 | 1854.19 |
| $P_{BESS}^{rated}$ [kW] | - | 1050 | 715.86 |
| Optimal BESS DoD [%] | - | 100 | 60.2 |
| BESS annual charge/discharge cycles | - | 755 | 369 |
| PV system rating [kW] | - | 300 | 300 |
| Monthly $P_{g,mo,s}^{max}$ for winter, spring, summer, and fall [kW] | 1081.31 | 445.76 | 288.11 |
| $P_{g,an}^{max}$ [kW] | 1081.31 | 445.76 | 288.11 |
| XFCS operation cost [$/year] | 356627.53 | 75751.12 | 181444.81 |
| BESS investment cost [$/year] | 0 | 136648.38 | 111112.76 |
| PV system investment cost [$/year] | 0 | 56562 | 56562 |
| Total demand charges [$/year] | 149220.57 | 61515.28 | 39759.07 |
| Total XFCS savings [$/year] | 0 | 175371.32 | 116969.36 |
| Total XFCS savings [%/year] | 0 | 34.66 | 23.12 |

The optimal PV sizing system is found to be 300 kW in both Case I and Case II, which is equal to the maximum allowed PV system size for a station with a total area of $2000 \, m^2$. Thus, it demonstrates that the installation of the PV system in the studied XFCS is economically viable in both cases because it serves multiple purposes by directly feeding the charging station demand, charging the BESS, and exporting the excess generated power to the PDN to earn extra revenue (see Figs. 9(b) and (f)). Note that in Table 3, the average power imported from the PDN in Base Case (i.e., 1081.31 kW) is higher than the maximum XFCS peak power demand at any time (i.e., 1050 kW), which is attributed to the efficiency losses in the power electronic converters.

Case I promises comparatively significant payoffs, but it requires 755 BESS annual charge/discharge cycles (more than 2 times as in Case II) with 100% DoD. Consequently, the BESS in Case I will reach its End-of-Life (EOL) in only ~3.7 years and will need to be replaced, thereby incurring extra investment costs. Conversely, the BESS undergoes only 369 annual cycles with a DoD of 60.2% in Case II and will not be replaced during the project's lifetime, thus preventing any additional investment costs. Note that, the EOL is estimated using $\frac{\text{maximum allowable BESS cycles based on the selected DoD}}{\text{BESS annual cycles of operation}}$ with help of the cycle-life curve presented in Fig. 5. Nevertheless, it should be noted that for the EOL estimation, it is assumed that BESS operation is strictly following the 60.2% DoD with annual 369 cycles of operation. The DoD restriction can be practically imposed by limiting BESS charge/discharge operation between the SoC range of 100%-39.8%.

In the light of results presented in Table 3 and discussion, this case study signifies that accounting for the BESS life degradation aspects is imperative and cannot be ignored in charging station planning studies.

For a typical weekend and weekday in the winter season (i.e., correspondingly scenarios #1 and #5), the optimal energy management of the XFCS, corresponding BESS SoC variations in response to station's charging demand and electricity price variations, and average power imported from the PDN are displayed in Fig. 9. For the purpose of discussion, this section presents the results for the winter season only, however, the proposed model can give the optimal energy management for the station and BESS operation for all studied seasons. In Figs. 9(b) and (f), the ellipses E1, E2, and E3 indicate the periods when BESS discharges to earn revenue by feeding the XFCS charging demand and/or exporting energy to the PDN during high electricity price periods. The PV system generates power to fulfill the station's charging demands partially/fully and either export the excess energy to the PDN or charge the BESS. During the weekend, the PV system generation coincides with the XFCS demand, therefore the majority of the time it is being utilized to feed the charging station demands. By contrast, as there is less charging demand on the station during the weekday daytime hours, the excess PV power is either exported to the PDN to earn revenue, or it charges the BESS for later use during peak price hours.

Charging/discharging of the BESS and corresponding SoC variations are illustrated in Figs. 9(c) and (g). The BESS operation is always constrained by the optimal DoD of 60.2% to ensure longevity. It is evident in both charts that BESS charges during the off-peak price hours and discharges during the peak-price hours to take advantage of the energy arbitrage opportunities and earn revenue. Consequently, it shifts the charging station demand from peak hours to off-peak hours, and it also helps reduce peak demand on the local distribution network. Hence, energy arbitrage using BESS is not only beneficial for the XFCS owner but can also play a central role in reducing the peak demand on the distribution network, thereby precluding the potential future grid reinforcements which may otherwise be needed if multiple XFCS are installed on the PDN.

The BESS also discharges to meet another important objective: keeping the maximum monthly and annual average power imports from the PDN to the lowest and reducing the ensuing demand charges, as exemplified by Figs. 9(c), (g), (d), and (h). From Table 3, the total demand charge comparisons between Base Case and Case II reveal a reduction of ~73% that is realized by optimally utilizing the BESS operation to bring the maximum monthly and annual average power imports from 1081.31 kW to 288.11 kW.

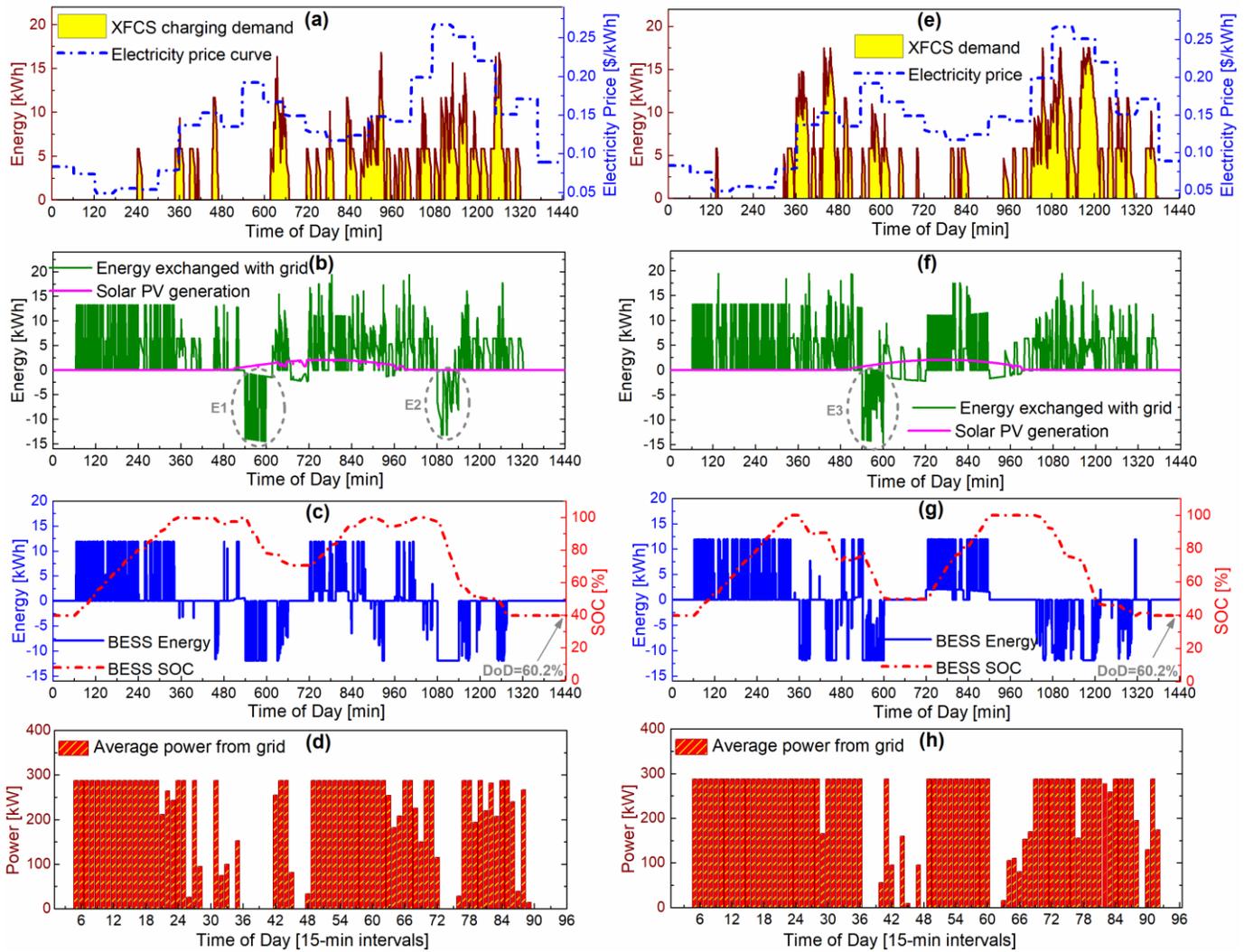

Fig. 9. Deterministic optimization model results with project lifetime of 20 years: winter season results for the weekend (scenario #1) and weekday (scenario #5) are respectively shown in (a) – (d) and (e) – (h). (a), (e) Hourly electricity price and the XFCS charging demand curves; (b), (f) energy exchanged with the PDN to satisfy the XFCS charging demand, to charge/discharge the BESS for arbitrage, and to send excess PV generation back to PDN to earn extra revenue. Positive values represent the energy imported from the PDN, and negative values indicate the energy exported back to the PDN; (c), (g) BESS SOC variations in percentage and energy flows in kWh: positive values represent energy to charge the ESS and negative values represent the energy discharged from the BESS; and (d), (h) average power imported from the PDN (averaged over 15-min time intervals)

Note that the efficacy of the proposed MILP formulations to tally the BESS cycles using cumulative charge/discharge energy is evident in Figs. 9(c) and (g), wherein the BESS charges and discharges frequently during a day—charging (/discharging) happens before finishing a discharge (/charge) cycle—and accurate cycle counting using approaches presented in the literature [2, 65, 66] may not be practical. Finally, Figs. 9(d) and (h) present the daily average power import from the PDN for both weekends and weekdays of the winter season, wherein the average power import is always either equal to or less than 288.11 kW. Hence, it exemplifies that the proposed model keeps the average grid power imports minimized, at all times, to keep the ensuing total demand charges to the lowest.

Table 4. Simulation results of XFCS sizing with different planning horizons

| Project lifetime [years] | Base Case (w/o PV & BESS) | $\mathcal{L}$=5 | $\mathcal{L}$=10 | $\mathcal{L}$=15 | $\mathcal{L}$=20 |
|---|---|---|---|---|---|
| $C_{BESS}$ [kWh] | - | 358.22 | 797 | 1394.03 | 1854.19 |
| $P_{BESS}^{rated}$ [kW] | - | 358.22 | 582.62 | 671.54 | 715.86 |
| Optimal BESS DoD [%] | - | 88.53 | 68.85 | 62.03 | 60.2 |
| BESS annual charge/discharge cycles | - | 710 | 582 | 468 | 369 |
| PV system rating [kW] | - | 0 | 300 | 300 | 300 |
| Monthly $P_{g,mo,s}^{max}$ for winter, spring, summer, and fall [kW] | 1081.31 | 684.39 | 435.74 | 337.22 | 288.11 |
| $P_{g,an}^{max}$ [kW] | 1081.31 | 684.39 | 435.74 | 337.22 | 288.11 |
| XFCS operation cost [$/year] | 356627.53 | 325064.25 | 213986.85 | 192001.10 | 181444.81 |
| BESS investment cost [$/year] | 0 | 80149.52 | 90202.68 | 105656.74 | 111112.76 |
| PV system investment cost [$/year] | 0 | 0 | 90519 | 67710 | 56562 |
| Total investment cost [$/year] | 0 | 80149.52 | 180721.68 | 173366.74 | 167674.76 |
| Total demand charges [$/year] | 149220.57 | 94445.21 | 60132.45 | 46535.98 | 39759.07 |
| Total XFCS savings [$/year] | 0 | 6189.10 | 51007.1 | 93944.10 | 116969.36 |
| Annualized return on investment (AROI) [%] | 0 | 7.72 | 28.22 | 54.18 | 69.75 |
| Total XFCS savings [%/year] | 0 | 1.22 | 10.08 | 18.57 | 23.12 |

### 5.3.2 *Impact of Planning Horizon on XFCS Sizing using Deterministic Optimization Model*

This section investigates the impacts of different project lifetimes and offers insights and discussion on the results listed in Table 4. Base Case represents the case without considering the application BESS and PV system in the XFCS. Results reveal that a higher rated BESS is economically more viable with a longer project lifetime; therefore, with $\mathcal{L}$=20 years, the energy and power ratings are highest among all studied cases. This is because the return on large investment is more promising and yields larger payoffs with longer project lifetimes, i.e., 23.12% savings are expected with $\mathcal{L}$=20 years versus only 1.22% in case of $\mathcal{L}$=5 years. Moreover, the maximum average power imported from the PDN and resultant demand charges are highest for $\mathcal{L}$=5 years and lowest for $\mathcal{L}$=20 years. This is because, with higher-rated BESS, there is more opportunity to utilize it for reducing the average power imported from the PDN.

Annualized return on investment (AROI) is a commonly used metric for the economic valuation of any investment and is defined as the ratio of annual savings over annualized investment [115-118]. Table 4 shows that with $\mathcal{L}$=20 years the AROI is maximum (i.e., 69.75%) and is minimum (i.e., 7.72%) with $\mathcal{L}$=5 years assuming constant operation of BESS and PV system with no reliability issues. The AROI values indicate promising returns on investment with longer project lifetimes. Note that, per [116], on average, investors expect after-tax AROI of about 30-40%. In this work, the large AROI values are chiefly attributed to the savings realized by using the BESS for energy arbitrage and reduction in total demand charges.

With longer project lifetimes, the optimal value of the BESS DoD is smaller; additionally, the BESS undergoes fewer annual cycles to increase its longevity and thereby ensuring that it will not reach its EOL before the end of the project. Among the studied cases, investment in the PV system is not economically feasible with $\mathcal{L}$=5 years, and for the rest of the cases, installing the maximum allowed PV capacity is proved valuable.

Based on the findings and discussion, it is deduced that higher investments in the PV system and BESS yields higher AROI and savings in the XFCS total annualized cost with longer project durations. Therefore, $\mathcal{L}$=20 years is used to perform the rest of the case studies for this paper.

### 5.3.3    XFCS Sizing using Robust Optimization Model

In the RO model, the uncertainty bounds of $U^\lambda = \pm 20$, $U^D = \pm 10$, $U^{PV} = \pm 20$ are considered to accommodate forecast errors in the electricity price, XFCS demand, and PV system generation. The following cases are analyzed to study the impact of uncertainties on XFCS sizing and annualized costs:

- Case III: Impact of uncertainty in the electricity market price on XFCS sizing

- Case IV: Simultaneous impact of all uncertainties on XFCS sizing

**Case III:** In this case study, the sensitivity of XFCS components sizing and the total cost is appraised with different levels of robustness against the electricity price signal. Fig. 10 illustrates the results when the robust parameter for price ($\Gamma^\lambda$) is varied from 0 to 100%. Where $\Gamma^\lambda$=0% represents the deterministic case and $\Gamma^\lambda$=100% is the most conservative solution, wherein the electricity price curve either picks lower or upper bounds from its uncertainty set to occasion a worst possible impact on the total cost. Note that the worst-case cannot be realized by simply using the upper bound of the electricity price uncertainty since it may help XFCS owner earn more revenue by exporting energy back to the PDN at higher prices. Therefore, the RO model picks the worst case such that it causes the worst possible impact on total cost and payoffs.

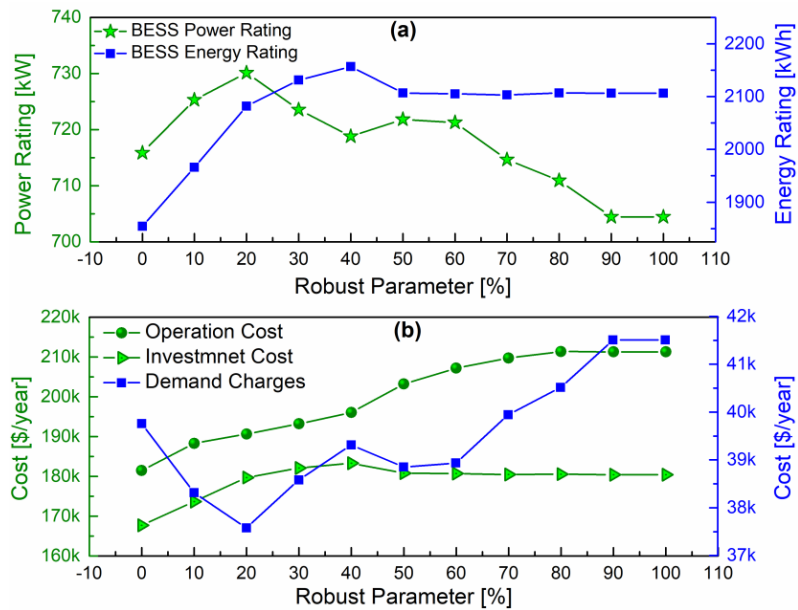

Fig. 10. Robust optimization model results: (a) sensitivity of BESS sizing and (b) XFCS cost components with different levels of robustness against uncertainty in forecasted electricity prices

The impact of increasing $\Gamma^\lambda$ on the sensitivity of power ratings for BESS is not substantial. In contrast, increasing the $\Gamma^\lambda$ causes a notable increase in BESS energy rating, which increased from 1854 kWh ($\Gamma^\lambda$=0%) to 2107 kWh ($\Gamma^\lambda$=100%). This increase in the BESS sizing is to hedge against the uncertainty of electricity prices: with an increase in the electricity market price (i.e., with increasing $\Gamma^\lambda$), feeding XFCS demand using energy import from the PDN becomes less cost-effective, therefore BESS energy ratings get increased to safeguard the XFCS operation against increased electricity prices and to exploit potential energy arbitrage opportunities. Fig. 9 shows that the increase in BESS sizing is more sensitive to the robust parameter at the start until $\Gamma^\lambda = 40\%$ than for higher values of $\Gamma^\lambda$. This is because the RO initially picks electricity price values for the time instants in which it would cause the worst impact on the XFCS operation cost. Therefore, BESS sizing increases aggressively to make the XFCS operation more economical during those periods. Since the PV system was already rated at its maximum allowable capacity, its rating remains unchanged. Increasing $\Gamma^\lambda$ yields increased investment cost, which is referred to as the *price of robustness* in the literature, and it is paid to hedge against the uncertainties [22]. Compared to the deterministic case, the increase in the total investment cost is 7.6% for the most risk-averse case.

From Fig. 10, note that variations in the demand charges are mirror images of changes in the BESS power ratings: with higher power rated BESS the annual demand charges are lower. Additionally, the highest and lowest demand charges respectively correspond to the lowest and highest power ratings for the BESS. This is because BESS with higher power ratings can help lower the maximum average power imported from the grid, and ensuing demand charges, more effectively than a BESS with smaller power ratings. Therefore, it is concluded that demand charges are directly correlated with BESS power rating when compared to energy capacity.

**Case IV:** This case study seeks to explore the impacts of different levels of robustness against all uncertainties. Different levels of robust parameters are simultaneously selected for all uncertainties and impacts on XFCS sizing and the station's cost components are analyzed.

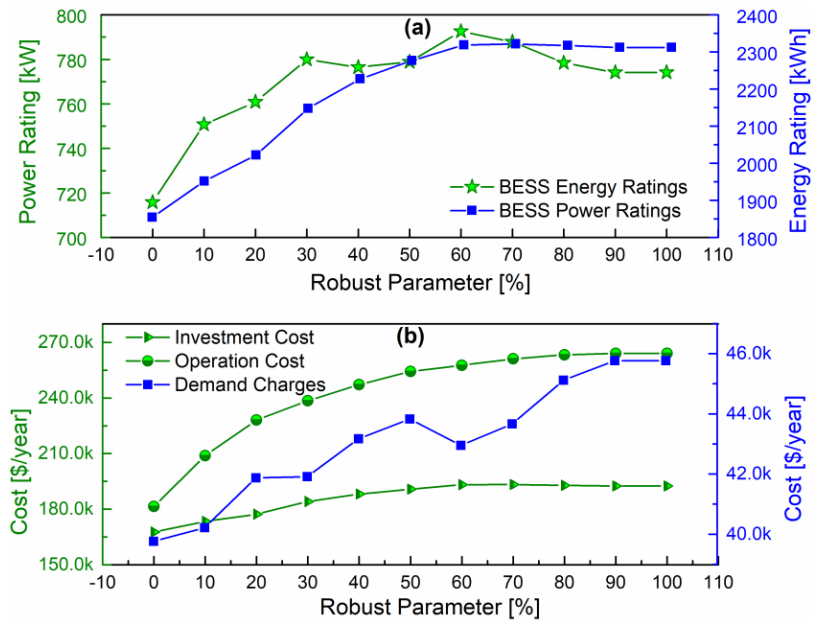

Fig. 11. Robust optimization model results: (a) sensitivity of BESS sizing and (b) XFCS cost components with different degrees of robustness against all uncertainties

Fig. 11 illustrates the effects of increasing robust parameter on BESS ratings and XFCS cost components. The power and energy ratings of the BESS respectively increased from 1854 kWh and 716 kW (deterministic case) to 2312 kWh and 774 kW (most risk-averse case). Consequently, the total yearly investment increases by ~15% for the most conservative case. In the deterministic case, the cost of investment is lower, but the XFCS is more vulnerable to uncertainties in the input parameters and may result in more operating costs. In the most conservative case, the investment cost becomes higher to secure the XFCS operation against uncertainties and their worst possible impact on the total cost of the station. Moreover, the PV system size remains unaffected because it was already equal to the maximum allowed ratings in the deterministic case. In addition, as mentioned in Case III, demand charges variations are strongly correlated with the power sizing of the BESS, as seen in Fig. 11.

Based on the results and discussion presented in Sections 5.3.1, 5.3.2, and 5.3.3, this paper recommends using a BESS of 1854 kWh/716 kW ratings and a solar system of 300 kW power rating for the deterministic case. For the robust case (the most risk-averse case), BESS is sized as 2312 kWh/774 kW and the solar system as 300 kW power rating. Note that the project lifetime of 20 years is considered for these recommendations since it yields the highest annualized savings and returns on investment among all considered project durations.

### 5.3.4    Sensitivity Analysis

This section studies the sensitivity of BESS and PV system sizing with input parameters, namely electricity market price and investment cost of the BESS and PV system.

*5.3.4.1 Sensitivity of Sizing with Electricity Price*

Electricity price is varied using the electricity price multiplier (EPM) and its impact on the sizing of XFCS components and annualized savings is studied. Note that the EPM=1.0 belongs to the deterministic case (i.e., Case II) that is solved with the original price signal, as displayed in Fig. 8. Fig. 12(a) illustrates that with EPM values of less than 0.6, the application of PV system in the studied station is not economically viable; this is because, with lower electricity prices, the return on investment in the PV system is insufficient to justify the investment. With EPM≥0.6, the PV system is rated at its maximum allowable capacity for the considered total area of the XFCS. This is because, with higher EPM, the PV system proves its worth by directly feeding the charging station demand, charging the BESS for later use during peak prices, and exporting the excess generated power to the PDN to earn extra revenue.

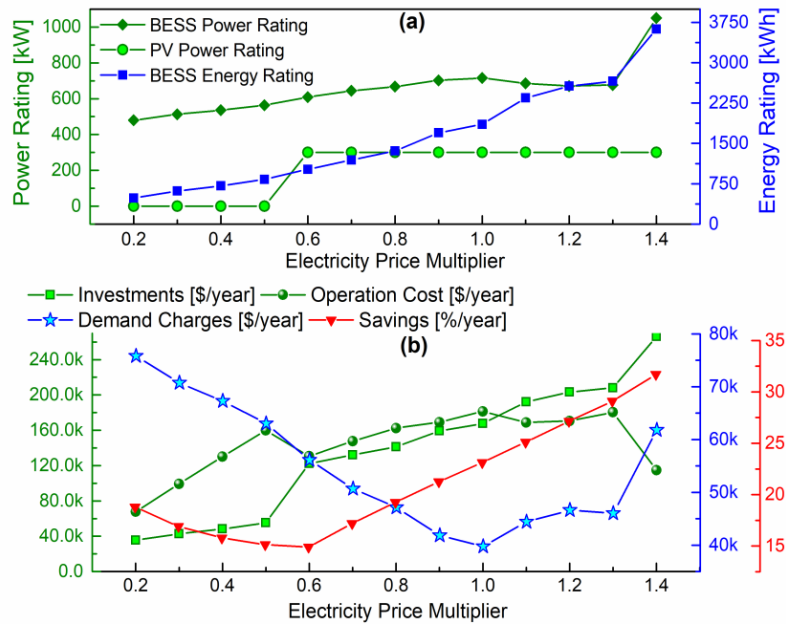

Fig. 12. Impact of varying EPM on (a) BESS and PV system sizing, (b) different cost components and total annualized savings

Note that with lower values of EPM, the investment in BESS is always feasible, albeit with smaller power and energy ratings. This is because for lower EPM values, the energy arbitrage using BESS may not be economically beneficial, but BESS still finds its application in lowering the demand charges. In Fig. 12 (b), the total investment cost curve shows an increasing trend with increasing values of EPM. This is attributed to the fact that investing in BESS for higher EPM values brings more revenue from energy arbitrage either by exporting the charged energy back to the PDN or utilizing it to feed the XFCS demand during peak-price hours. A sudden increase in the total investment cost curve and a sudden decrease in the operation cost curve is observed in Fig. 12(b) when EPM increases from 0.5 to 0.6, which is because of deploying a PV system with the rated power of 300 kW.

In Fig. 12(b), the demand charges curve initially decreases consistently until EPM=1.0, while for EPM>1.0 it shows an increasing trend with a sudden increase at the end. This is because, with EPM>1.0, BESS finds more economic benefits of its usage in energy arbitrage than lowering the demand charges. Hence with EPM=1.4, there is a sudden decrease in the operation cost and a sudden increase in the demand charges. Finally, the saving curve initially decreases until EPM=0.6 and then it increases consistently afterward. The initial decrease in annual savings is attributed to the fact that with EPM<0.6, the energy arbitrage and PV deployment is not economically feasible, and with an increase in EPM from 0.2 to 0.6, the energy purchased from the PDN becomes more costly, thereby making the saving curve to exhibit a decreasing trend. However, for EPM≥0.6 total cost of the station becomes lower, and savings consistently increase mainly because of exporting excess solar generation back to the PDN to earn extra revenue and exploiting the energy arbitrage prospects using BESS. Hence, reduction in the operation cost and demand charges overshadows the investment costs of BESS and PV systems with higher EPM values and results in higher annualized savings.

### 5.3.4.2   Sensitivity of Sizing with Investment Cost

The investment cost of both BESS and PV systems is varied using investment cost multiplier (ICM) and its effect on sizing and annualized saving is explored. With ICM=1.0, results are the same as presented for Case II of Section 5.3.1, which was solved with original investment cost data. Fig. 13 depicts the impact of changing ICM on BESS and PV system sizing, XFCS cost components, and total savings. From Fig. 13(a), with an increase in ICM both power and energy ratings of the BESS exhibit a consistent decrease, while PV system size remains the same until ICM=1.6. For higher ICM values, it becomes economically infeasible to deploy a PV system of any rating. Consequently, a sudden increase in the operation cost and a sudden decrease in the investment cost is evident from Fig. 13(b) when ICM increases from 1.6 to 1.7.

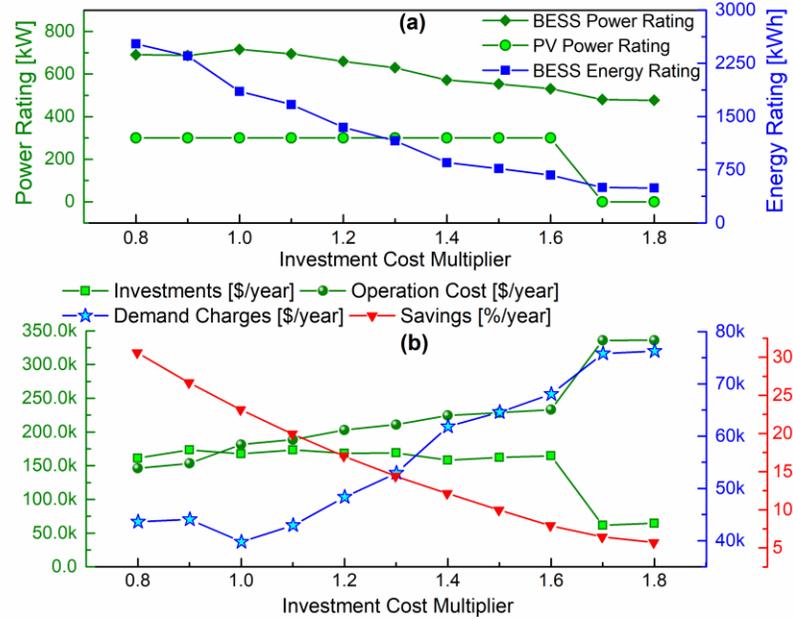

Fig. 13. Impact of varying ICM on (a) BESS and PV system sizing, (b) different cost components and total annualized savings

Note that investing in BESS remains economically viable for higher ICM, mainly because it can prove its value by lowering the demand charges even with ICM=1.8. The demand charges curve in Fig. 13 (b) shows an increasing tendency as ICM values increase. This is because, with increasing ICM values, the power rating of the BESS keeps decreasing and therefore results in higher demand charges. Lastly, the saving curve exhibits a decreasing trend with an increase in ICM values which can be attributed to the fact that with higher ICM values, investing in large BESS and PV systems becomes costly and may not justify returns via energy arbitrage at the expense of bigger investment cost. Hence, it limits the ability of the BESS to exploit the potential energy arbitrage opportunities and the resultant revenue, thus resulting in decreased annualized savings.

*5.3.5    Sensitivity of XFCS Sizing with EV Users' Departure Times and Number of XFCS Charging Ports*

This section investigates the sensitivity of EV users' departure times from their parking places and the number of charging ports at the XFCS on the sizing of the BESS and PV system. The $EVC_1$ category of EVs has the largest share of 61% in the studied XFCS demand model and influences the XFCS demand profile. EVs in this category follow a normal probability distribution for departure time from their parking places. In this analysis, the sensitivity of the XFCS components' sizing with mean values of departure time in $\mathcal{N}_2(06:52, 01:18)$ is studied for EVs in category $EVC_1$. EVs in categories $EVC_2$ and $EVC_3$ are assumed to follow the same $\mathcal{N}_1(13:51, 5:12)$ distribution for their departure. Fig. 14 illustrates that the variations in EVs' mean departure time have a significant impact on BESS energy ratings, moderate effect on BESS power ratings, and no effect on PV system ratings. With an increase in mean departure time, BESS energy capacity increases and achieves the highest value with a mean departure time of 08:52. After which, it keeps decreasing and achieves the lowest rating with a mean departure time of 13:52. This is attributed to the fact that variations in EVs' mean departure time affect the XFCS demand peaks which in turn affect the capability of BESS to participate in energy arbitrage opportunities. If the XFCS demand profile peaks coincide with the electricity price curve, the BESS can discharge during the peak-price demand hours and avoid the high cost of XFCS operation. Hence, BESS sizing gets increased in this case. On the contrary, if EVs mean departure time shift the XFCS demand peaks to off-peak hours, there may be fewer energy arbitrage opportunities and BESS sizing will be decreased, as illustrated in Fig. 14. The BESS energy ratings are lowest with EVs' mean departure times of $03:52$ and $13:52$. Based on the sizing results in Fig. 14, the correlation of BESS power ratings with EVs' mean departure time is unclear. The PV system was sized at maximum allowable ratings and exhibited no change with EVs' mean departure time variations.

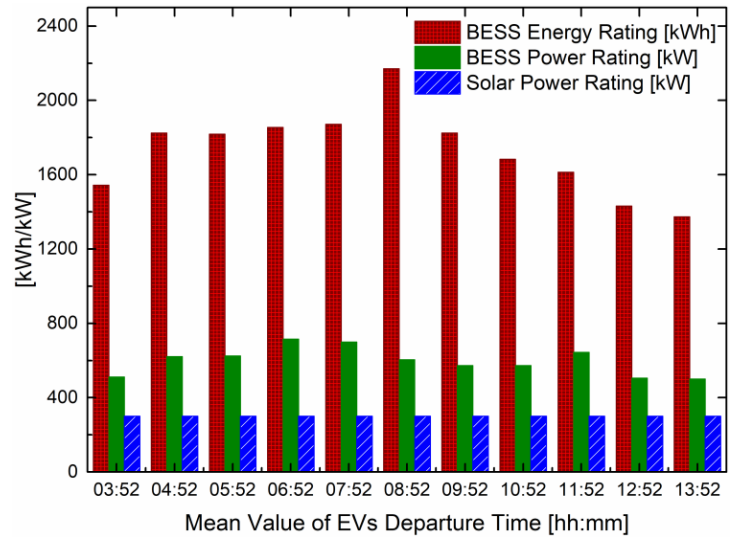

Fig. 14. Impact of varying EVs' mean departure time on BESS and PV system sizing

The sensitivity of BESS and PV system sizing with the number of XFCS charging ports in analyzed in another study and results are shown in Fig. 15. Note that the demand profiles used in the rest of the paper are obtained with $r = 3$ charging ports and $w = 5$ waiting spots. For this analysis, waiting spots are kept the same and only the number of charging ports are changed. With the increasing number of charging ports, BESS power and energy ratings increase. This can be attributed to the fact that with a higher number of XFCS ports, more EVs can get recharged simultaneously at the station without staying in the waiting zone, which would result in higher demand peaks. Moreover, EVs may not need to wait, thus resulting in less rejection rates of their charging requests. In contrast, with fewer charging ports, more EVs will have to stay in the waiting zones until a port becomes available, and there is a higher possibility of rejection of EV's charging request, thus decreasing the demand peaks and total area under the charging demand curve. Lastly, the PV system rating remains unaffected, and it is sized at its maximum allowable capacity.

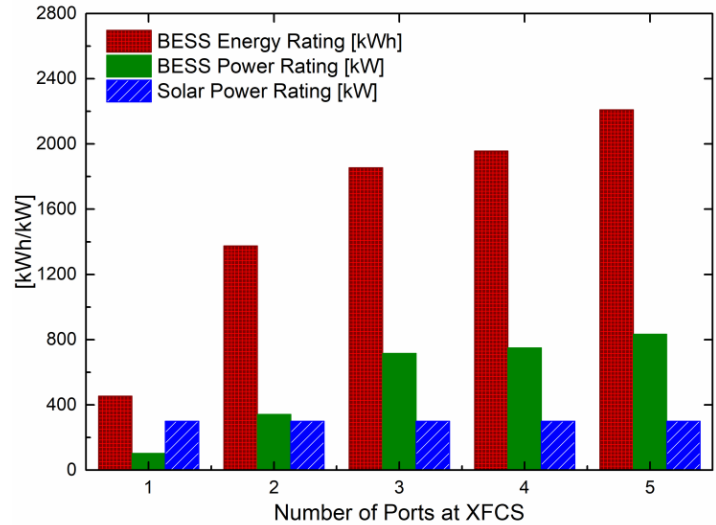

Fig. 15. Impact of varying the number of XFCS charging ports on the BESS and PV system sizing

### 5.3.6    Discussion on McCormick Relaxations

As described in Section 3, McCormick relaxations are used to relax the bi-linear terms appearing in (36) and (37), and consequently, linear formulations are obtained. Relaxing the original NLP model using McCormick approximation helps reduce the computational intricacy of the original model at the expense of solution's quality; the final optimal solution obtained from the relaxed model may not always be the exact optimum of the original NLP model, instead, it provides an optimistic bound for the optimal solution of the original NLP model. With tighter lower and upper bounds of variables involved in the bi-linear terms, the quality of the relaxed solution gets improved [92]. Practically, solution quality obtained from the relaxed model is appraised by solving the original NLP model using the obtained solution and analyzing the feasibility of the constraints. If all the constraints are not satisfied, then the solution obtained from the relaxed model is not feasible for the original NLP model [119].

In this work, both bi-linear terms, appearing in (36) and (37), are associated with the operational constraints of the BESS, therefore, to appraise the quality of the relaxed solution, the planning decisions, obtained for Case II (see Table 3), were fixed and used as parameters in the original NLP model and it was re-solved. The $C_{BESS}$ appears in both bi-linear terms of (36) and (37) in the original NLP model, therefore, fixing $C_{BESS}$ and using it as a parameter makes both bi-linear terms vanish, thereby rendering the original NLP model a linear model that is easily solvable.

Table 5. Quality of solution obtained from the relaxed model

|  | Solution I: relaxed model | Solution II: original model with fixed planning decisions from Solution I | Gap |
|---|---|---|---|
| Investment cost [$/year] | 167674.76 | 167674.76 | 0% |
| Operation cost [$/year] | 181444.81 | 181446.64 | 0.001% |
| Total demand charges | 39759.07 | 39759.43 | 0% |

| | | |
|---|---|---|
| [$/year] Total XFCS savings [$/year] | 116969.36 | 116967.20 | 0.0018% |

Table 5 presents a comparison of the results by following the procedure explained in the previous paragraph. The investment cost in both solutions is the same because the investment decisions for Solution I were used to obtain Solution II. The percent gap between both solutions is insignificant, demonstrating the efficacy of using tighter lower and upper bounds of the variables involved in the bi-linear terms of the original NLP problem.

It can be inferred that the proposed relaxed linear model is a good approximation of the original NLP model and can be utilized to carry out charging station planning studies.

## 6    Conclusions

This paper proposed a MILP-based optimization model to obtain BESS and PV system sizing and optimal energy management for an XFCS, distinctively for each representative scenario of a typical year. The BESS was sized to serve four vital purposes: i) to satisfy the XFCS demand during peak price hours, ii) to store excessive PV system generation, iii) to exploit the energy arbitrage opportunities by charging during low price hours and discharging to export energy back to the PDN to maximize net revenue, and iv) to manage the monthly and annual average power imports from the PDN to minimize total demand charges. Case studies demonstrated the proposed model's efficacy. It was inferred that by considering the BESS life degradation, it will not reach the EOL during the project lifetime, thus preventing additional investment costs. Case study results implied that the BESS would need to be replaced after ~3.7 years if degradation aspects are not considered in the XFCS sizing studies. In addition, the effectiveness of the proposed MILP formulations to tally the BESS cycles was demonstrated when the BESS charged and discharged discontinuously during a day, under which conditions the cycle counting approaches presented in the literature may not be practical.

Furthermore, a reduction of ~73% in total demand charges was realized by optimally utilizing the BESS operation which reduced the maximum monthly and annual average grid power imports from 1081.31 kW to 288.11 kW. Case studies signified that the highest annualized savings of 23.12% and AROI of 69.75% were expected when the XFCS components were sized for a planning horizon of 20 years, and savings of only 1.22% and AROI of 7.72% were achieved with a planning horizon of 5 years. An optimal PV system rating was equal to the maximum allowable size for a station with the considered area of $2000 \ m^2$. Sensitivity analyses were performed to provide insights into how changing input parameters and different levels of robustness against uncertainties in the input data impact sizing of the BESS, the PV system, and the total cost of the XFCS. It is found that demand charges are strongly correlated with BESS power rating when compared to its energy rating. Moreover, with EPM values of less than 0.6, the application of the PV system in the studied station is not economically viable, while it is rated at its maximum allowable capacity for EPM≥0.6. The BESS, on the other hand, still finds its application in lowering the demand charges even with EPM=0.2. In addition, it becomes economically infeasible to deploy a PV system of any rating with ICM≥1.6, while the BESS remains economically viable even with ICM=1.8 because it can prove its worth in demand charges reduction.

Finally, the quality of the solution obtained from the relaxed model was appraised, and the proposed relaxed model was found to be a good approximation of the original NLP model, warranting its application for planning studies of charging stations.

## Appendix

*Tight Piece-wise McCormick Relaxation for Bi-linear Terms: MILP Formulations*

This section presents MILP formulations to obtain a tight piece-wise McCormick relaxation for bi-linear term in (38) with a bi-variate partitioning approach. Assuming $\mathscr{g}_{xnyn'}$ be the binary variable representing the active partitioning for $\Psi$ and $C_{BESS}$, which are respectively bounded by $[\underline{\Psi}, \overline{\Psi}]$ and $[\underline{C_{BESS}}, \overline{C_{BESS}}]$. The piece-wise McCormick relaxation for $\Upsilon = \Psi \times C_{BESS}$ with bi-variate partitioning can be characterized as Generalized Disjunctive Program (GDP), as given by (A.1):

$$\bigvee_{n=1}^{\mathbb{N}} \bigvee_{n'=1}^{\mathbb{N}} \begin{bmatrix} \mathscr{g}_{xnyn'} \\ \Upsilon_{xy} \geq \Psi_x . \underline{C_{BESS\,yn'}} + \underline{\Psi_{xn}} . C_{BESS\,y} - \underline{\Psi_{xn}} . \underline{C_{BESS\,yn'}} \\ \Upsilon_{xy} \geq \Psi_x . \overline{C_{BESS\,yn'}} + \overline{\Psi_{xn}} . C_{BESS\,y} - \overline{\Psi_{xn}} . \overline{C_{BESS\,yn'}} \\ \Upsilon_{xy} \leq \Psi_x . \overline{C_{BESS\,yn'}} + \underline{\Psi_{xn}} . C_{BESS\,y} - \underline{\Psi_{xn}} . \overline{C_{BESS\,yn'}} \\ \Upsilon_{xy} \leq \Psi_x . \underline{C_{BESS\,yn'}} + \overline{\Psi_{xn}} . C_{BESS\,y} - \overline{\Psi_{xn}} . \underline{C_{BESS\,yn'}} \\ \underline{\Psi_{xn}} \leq \Psi_x \leq \overline{\Psi_{xn}} \\ \underline{C_{BESS\,yn'}} \leq C_{BESS\,y} \leq \overline{C_{BESS\,yn'}} \end{bmatrix}$$

$$\forall (x,y) \in BL \text{ and } \mathscr{g}_{xnyn'} \in \{0,1\}, \forall \{x | (x,y) \in BL\}, n \in \{1, \dots, \mathbb{N}\}, n' \in \{1, \dots, \mathbb{N}\} \quad (A.1)$$

where $BL$ is an $(x,y)$-index set defining the bi-linear term $\Upsilon$ in (38).

$$\left. \begin{matrix} \underline{\Psi_{xn}} = \underline{\Psi_x} + \frac{(n-1).\left(\overline{\Psi_x} - \underline{\Psi_x}\right)}{\mathbb{N}} \\ \overline{\Psi_{xn}} = \underline{\Psi_x} + \frac{n.\left(\overline{\Psi_x} - \underline{\Psi_x}\right)}{\mathbb{N}} \end{matrix} \right\}, \; \forall \{x | (x,y) \in BL\}, n \in \{1, \dots, \mathbb{N}\} \quad (A.2)$$

$$\left. \begin{matrix} \underline{C_{BESS\,yn'}} = \underline{C_{BESS\,y}} + \frac{(n-1).\left(\overline{C_{BESS\,y}} - \underline{C_{BESS\,y}}\right)}{\mathbb{N}} \\ \overline{C_{BESS\,yn'}} = \underline{C_{BESS\,y}} + \frac{n.\left(\overline{C_{BESS\,y}} - \underline{C_{BESS\,y}}\right)}{\mathbb{N}} \end{matrix} \right\}, \; \forall \{x | (x,y) \in BL\}, n' \in \{1, \dots, \mathbb{N}\} \quad (A.3)$$

Applying the convex hull formulations to transform the linear GDP of (A.1) into MILP formulations, as given by (A.4) and (A.5):

$$\Upsilon_{xy} \geq \sum_{n=1}^{\mathbb{N}} \sum_{n'}^{\mathbb{N}} \left( \widehat{\Psi_{xxnyn'}} \cdot C_{BESS_{yn'}} + \underline{\Psi_{xn}} \cdot \widehat{C_{BESS}_{yxnyn'}} - \underline{\Psi_{xn}} \cdot C_{BESS_{yn'}} \cdot \mathscr{g}_{xnyn'} \right)$$

$$\Upsilon_{xy} \geq \sum_{n=1}^{\mathbb{N}} \sum_{n'}^{\mathbb{N}} \left( \widehat{\Psi_{xxnyn'}} \cdot \overline{C_{BESS_{yn'}}} + \overline{\Psi_{xn}} \cdot \widehat{C_{BESS}_{yxnyn'}} - \overline{\Psi_{xn}} \cdot \overline{C_{BESS_{yn'}}} \cdot \mathscr{g}_{xnyn'} \right)$$

$$\Upsilon_{xy} \leq \sum_{n=1}^{\mathbb{N}} \sum_{n'}^{\mathbb{N}} \left( \widehat{\Psi_{xxnyn'}} \cdot \overline{C_{BESS_{yn'}}} + \underline{\Psi_{xn}} \cdot \widehat{C_{BESS}_{yxnyn'}} - \underline{\Psi_{xn}} \cdot \overline{C_{BESS_{yn'}}} \cdot \mathscr{g}_{xnyn'} \right)$$

$$\Upsilon_{xy} \leq \sum_{n=1}^{\mathbb{N}} \sum_{n'}^{\mathbb{N}} \left( \widehat{\Psi_{xxnyn'}} \cdot C_{BESS_{yn'}} + \overline{\Psi_{xn}} \cdot \widehat{C_{BESS}_{yxnyn'}} - \overline{\Psi_{xn}} \cdot C_{BESS_{yn'}} \cdot \mathscr{g}_{xnyn'} \right) \Bigg\}, \forall (x,y) \in BL \quad \text{(A.4)}$$

$$\Psi_x = \sum_{n=1}^{\mathbb{N}} \sum_{n'}^{\mathbb{N}} \widehat{\Psi_{xxnyn'}}$$

$$C_{BESS_y} = \sum_{n=1}^{\mathbb{N}} \sum_{n'}^{\mathbb{N}} \widehat{C_{BESS}_{yxnyn'}}$$

$$\sum_{n=1}^{\mathbb{N}} \sum_{n'}^{\mathbb{N}} \mathscr{g}_{xnyn'} = 1$$

where $\underline{\Psi_{xn}}, \overline{\Psi_{xn}}, C_{BESS_{yn'}}, \overline{C_{BESS_{yn'}}}$ are given in (A.2) and (A.3), and

$$\underline{\Psi_{xn}} \cdot \mathscr{g}_{xnyn'} \leq \widehat{\Psi_{xxnyn'}} \leq \overline{\Psi_{xn}} \cdot \mathscr{g}_{xnyn'}$$

$$C_{BESS_{yn'}} \cdot \mathscr{g}_{xnyn'} \leq \widehat{C_{BESS}_{yxnyn'}} \leq \overline{C_{BESS_{yn'}}} \cdot \cdot \mathscr{g}_{xnyn'} \Bigg\} \forall \{x | (x,y) \in BL\}, n \in \{1, \dots, \mathbb{N}\}, n' \in \{1, \dots, \mathbb{N}\} \text{ and}$$

$$\mathscr{g}_{xnyn'} \in \{0,1\}, \forall \{x | (x,y) \in BL\}, n \in \{1, \dots, \mathbb{N}\}, n' \in \{1, \dots, \mathbb{N}\} \quad \text{(A.5)}$$

## Acknowledgment

This work was supported by the U.S. Department of Energy, under Grant DE-EE0008449. We thank Kirk Stetzel (LG Energy Solutions, Inc.) for providing support and insightful discussion on CPCV based extreme fast charging of EVs that improved the accuracy of XFCS daily demand modeling. We also thank Rodney Hilburn (Technology Applications Center (TAC), Ameren Illinois) and Mark Hagge (previously with TAC, Ameren Illinois) for providing invaluable information and discussions on utility practices and tariffs in Ameren.